\documentclass[]{llncs}

\usepackage{amssymb,amsmath}
\usepackage[table,dvipsnames]{xcolor}  
\usepackage{xfrac,xspace}
\usepackage{graphicx}
\usepackage{subcaption}  
\usepackage{wrapfig}
\usepackage{marginnote}
\usepackage[normalem]{ulem}  
\usepackage{tikz}
\usepackage{eurosym}
\usepackage[super]{nth}  
\usepackage{longtable} 
\usepackage{breakcites}
\usepackage[breaklinks  = true,
            colorlinks  = true,
            linkcolor   = black,
            urlcolor    = blue!50!green,
            citecolor   = Blue!50!blue,
            anchorcolor = Green]{hyperref}
\usepackage[nameinlink]{cleveref}
\usepackage{etoolbox}
\usepackage{lscape}
\usepackage[inline,shortlabels]{enumitem}
\usepackage{mathabx}
\usepackage{rotating}

\ifdefined \VersionLong
	\newcommand{\LongVersion}[1]{{\color{red!60!black}#1}}
	\newcommand{\ShortVersion}[1]{\ifdefined\VersionWithComments{\color{black!40}#1}\fi}
\else
	\newcommand{\LongVersion}[1]{\ifdefined\VersionWithComments{\color{red!60!black}#1}\fi}
	\newcommand{\ShortVersion}[1]{\ifdefined\VersionWithComments{\color{green!60!black}#1}\else#1\fi}
\fi

\graphicspath{{figures/},{imgs/}}  

\usetikzlibrary{decorations,fit,arrows,automata,positioning,shapes,shapes.multipart,calc,matrix,backgrounds,shapes.gates.logic,shapes.gates.logic.US,circuits}
\tikzstyle{every node}=[initial text=]
\tikzstyle{final}=[double]
\tikzstyle{sync}=[draw=blue,thick]
\tikzstyle{seq}=[path picture={
	\draw[->]  ([xshift=#1, yshift=6pt]path picture bounding box.south)
					-- ([xshift=#1, yshift=-7pt]path picture bounding box.south);}]

\newcommand{\leafAgent}[4][(0,0)]{%
	\node[state,initial] (l0) at #1 {$\loc_0$};
	\node[state,right of=l0] (l1) {$\loc_1$};
	\node[state,below of=l1, node distance=1cm] (l'1) {$\loc'_1$};
	\draw (l0) -- node[above] (action) {$#2$} (l1);
	\ifstrempty{#4}{}{
		\node[above of=action,node distance=0.4cm,
			text=PineGreen] (time) {$#4$};
	}
	\ifstrempty{#3}{}{
		\node[above of=time,node distance=0.4cm,
			text=PineGreen] (cost) {$#3$};
	}
	\draw[sync] (l1) to [loop right]
		node[above,yshift=0.1cm] {$\styleSync{\send{}#2\_\ok}$} (l1);
	\draw[sync] (l0) --
		node[below,sloped] {$\styleSync{\send{}#2\_\nok}$} (l'1);
	\draw[sync] (l'1) to [loop right]
		node[below,yshift=-1.5mm,pos=0.5] {$\styleSync{\send{}#2\_\nok}$} (l'1);
}

\newcommand{\andAgent}[6][(0,0)]{%
	\node[state,initial] (l0) at #1 {$\loc_0$};
	\node[state,right of=l0] (l1) {$\loc_1$};
	\node[state,right of=l1] (l2) {$\loc_2$};
	\node[state,right of=l2] (l4) {$\loc_{#2}$};
	\node[state,below of=l1] (l'1) {$\loc'_1$};

	\draw[sync] (l0) --
		node[above] {$\styleSync{\receive{}#3\_\ok}$} (l1);
	\draw[sync] (l1) --
		node[above] {$\styleSync{\receive{}#4\_\ok}$} (l2);

	\draw (l2) -- node[above] (action) {$#2$} (l4);
	\ifstrempty{#6}{}{
		\node[above of=action,node distance=0.4cm,
			text=PineGreen] (time) {$#6$};
	}
	\ifstrempty{#5}{}{
		\node[above of=time,node distance=0.4cm,
			text=PineGreen] (cost) {$#5$};
	}
	\draw[sync] (l4) to [loop right]
		node[above,yshift=0.1cm] {$\styleSync{\send{}#2\_\ok}$} (l4);

	\draw[sync] (l0) --
		node[above,sloped] {$\styleSync{\receive{}#3\_\nok}$} (l'1);
	\draw[sync] (l0) to [bend right=30]
		node[above,sloped] {$\styleSync{\receive{}#4\_\nok}$} (l'1);

	\draw[sync] (l'1) to [loop right]
		node[below,yshift=-1.5mm,pos=0.5] {$\styleSync{\send{}#2\_\nok}$} (l'1);
}

\newcommand{\orAgent}[8][(0,0)]{%
	\node[state,initial] (l0) at #1 {$\loc_0$};
	\node[state,right of=l0] (l1) {$\loc_1$};
	\node[state,right of=l1] (l2) {$\loc_{#2}$};
	\node[state,below of=l1] (l'1) {$\loc'_1$};
	\node[state,right of=l'1] (l'2) {$\loc'_2$};

	\draw[sync] (l0) to [bend left]
		node[above] (action1) {$\styleSync{\receive{}#3\_\ok}$} (l1);
	\ifstrempty{#6}{}{
		\node[above of=action1,node distance=0.4cm,
			text=PineGreen] (time1) {$#6$};
	}
	\ifstrempty{#5}{}{
		\node[above of=time1,node distance=0.4cm,
			text=PineGreen] (cost1) {$#5$};
	}
	\draw[sync] (l0) to [bend right]
		node[below] (action2) {$\styleSync{\receive{}#4\_\ok}$} (l1);
	\ifstrempty{#8}{}{
		\node[below of=action2,node distance=0.4cm,
			text=PineGreen] (time2) {$#8$};
	}
	\ifstrempty{#7}{}{
		\node[below of=time2,node distance=0.4cm,
			text=PineGreen] (cost2) {$#7$};
	}
	\draw (l1) -- node[above] {$#2$} (l2);
	\draw[sync] (l2) to [loop right]
		node[above,yshift=0.1cm] {$\styleSync{\send{}#2\_\ok}$} (l2);

	\draw[sync] (l0) to [bend right=30] node[below,sloped,very near end]
		{$\styleSync{\receive{}#3\_\nok}$} (l'1);
	\draw[sync] (l'1) -- node[above]
		{$\styleSync{\receive{}#4\_\nok}$} (l'2);
	\draw[sync] (l'2) to [loop right]
		node[below,yshift=-1.5mm,pos=0.5] {$\styleSync{\send{}#2\_\nok}$} (l'2);
}

\newcommand{\nandAgent}[7][(0,0)]{%
	\node[state,initial] (l0) at #1 {$\loc_0$};
	\node[state,right of=l0] (l1) {$\loc_1$};
	\node[state,below of=l1] (l'1) {$\loc'_1$};
	\node[state,right of=l1] (l2) {$\loc_2$};
	\node[state,right of=l2] (lA) {$\loc_{#2}$};

	\draw[sync] (l0) --
		node[above] {$\styleSync{\receive{}#3\_\ok}$} (l1);
	\draw[sync] (l1) --
		node[above] {$\styleSync{\receive{}#4\_\nok}$} (l2);

	\draw[sync] (l0) -- node[above,sloped]
		{$\styleSync{\receive{}#3\_\nok}$} (l'1);
	\draw[sync] (l0) to [bend right] node[below,sloped]
		{$\styleSync{\receive{}#4\_\ok}$} (l'1);
	\draw (l2) -- node[above] (action) {$#2$} (lA);
	\ifstrempty{#5}{}{
		\node[below of=action,node distance=0.8cm,
			text=PineGreen] (cost) {$#5$};
	}
	\ifstrempty{#6}{}{
		\node[below of=cost,node distance=0.4cm,
			text=PineGreen] (time) {$#6$};
	}
	\ifstrempty{#7}{}{
		\node[above of=action,node distance=0.4cm,
			text=RedOrange] (condition) {$#7$};
	}
	\draw[sync] (lA) to [loop right]
		node[above,yshift=0.1cm] {$\styleSync{\send{}#2\_\ok}$} (lA);
	\draw[sync] (l'1) to [loop right]
		node[below,yshift=-1.5mm,pos=0.5] {$\styleSync{\send{}#2\_\nok}$} (l'1);
}

\newcommand{\sandAgent}[6][(0,0)]{%
	\node[state,initial] (l0) at #1 {$\loc_0$};
	\node[state,right of=l0] (l1) {$\loc_1$};
	\node[state,below of=l1,node distance=2cm] (l'1) {$\loc'_1$};
	\node[state,right of=l1] (l2) {$\loc_2$};
	\node[state,right of=l2] (lA) {$\loc_{#2}$};

	\draw[sync] (l0) --
		node[above] {$\styleSync{\receive{}#3\_\ok}$} (l1);
	\draw[sync] (l1) --
		node[above] {$\styleSync{\receive{}#4\_\ok}$} (l2);
	\draw[sync] (l0) -- node[above,sloped]
		{$\styleSync{\receive{}#3\_\nok}$} (l'1);
	\draw[sync] (l1) to node[above,sloped]
		{$\styleSync{\receive{}#4\_\nok}$} (l'1);
	\draw (l2) -- node[above] (action) {$#2$} (lA);
	\ifstrempty{#6}{}{
		\node[above of=action,node distance=0.4cm,
			text=PineGreen] (time) {$#6$};
	}
	\ifstrempty{#5}{}{
		\node[above of=time,node distance=0.4cm,
			text=PineGreen] (cost) {$#5$};
	}
	\draw[sync] (lA) to [loop right]
		node[above,yshift=0.1cm] {$\styleSync{\send{}#2\_\ok}$} (lA);
	\draw[sync] (l'1) to [loop right]
		node[below,yshift=-1.5mm,pos=0.5] {$\styleSync{\send{}#2\_\nok}$} (l'1);
}

\ifdefined \VersionWithComments
  \newcommand{\ceb}[1]{\textcolor{PineGreen}{\marginX{}[\textbf{Carlos}: #1 ]}}
  \newcommand{\ea}[1]{\textcolor{purple}{\marginX{}[\textbf{Étienne}: #1]}}
  \newcommand{\ja}[1]{\textcolor{WildStrawberry}{\marginX{}[\textbf{Jaime}: #1 ]}}
  \newcommand{\mk}[1]{\textcolor{magenta}{\marginX{}[\textbf{Michał}: #1]}}
  \newcommand{\ms}[1]{\textcolor{magenta!70!black}{\marginX{}[\textbf{Mariëlle}: #1
  ]}}
  \newcommand{\lp}[1]{\textcolor{green!70!black}{\marginX{}[\textbf{Laure}: #1 ]}}
  \newcommand{\wop}[1]{\textcolor{cyan!75!black}{\marginX{}[\textbf{Wojciech}: #1 ]}}
  
\else
  \newcommand{\ceb}[1]{}
  \newcommand{\ea}[1]{}
  \newcommand{\ja}[1]{}
  \newcommand{\mk}[1]{}
  \newcommand{\ms}[1]{}
  \newcommand{\lp}[1]{}
  \newcommand{\wop}[1]{}
\fi

\newcommand{\eg}{e.g.\xspace}
\newcommand{\ie}{i.e.\xspace}
\newcommand{\st}{s.t.\xspace}

\newcommand{\actAttack}{\ensuremath{a}}
\newcommand{\ActAttack}{\ensuremath{A}}
\newcommand{\actAttackSet}{\ensuremath{\Sigma_a}}
\newcommand{\actDefence}{\ensuremath{d}}
\newcommand{\ActDefence}{\ensuremath{D}}
\newcommand{\actDefenceSet}{\ensuremath{\Sigma_d}}
\newcommand{\gate}[1]{\ensuremath{\mathtt{\MakeUppercase{#1}}}\xspace}
\newcommand{\gateAND}{\gate{and}}
\newcommand{\gateOR}{\gate{or}}
\newcommand{\gateNAND}{\gate{nand}}

\newcommand{\gateSAND}{\gate{sand}}

\newcommand{\gateSNAND}{\gate{snand}}

\newcommand{\initTime}[1]{\ensuremath{\mathit{init\_time}(#1)}}
\newcommand{\Time}[1]{\ensuremath{\mathit{time}(#1)}}
\renewcommand{\EUR}[1]{\euro\,{\ensuremath{\mathrm{#1}}}\xspace}

\def\ADT/{ADT}
\def\AMAS/{AMAS}
\def\EAMAS/{EAMAS}
\newcommand{\loc}{\ensuremath{l}}
\def\IIS/{IIS}
\def\EIIS/{EIIS}
\def\MAS/{MAS}
\newcommand{\nok}{\ensuremath{\mathit{nok}}}
\newcommand{\ok}{\ensuremath{\mathit{ok}}}
\newcommand{\receive}{\ensuremath{?}}
\def\resp/{resp.}
\newcommand{\send}{\ensuremath{!}}
\newcommand{\styleSync}[1]{\textcolor{blue}{\ensuremath{#1}}}
\newcommand{\casestudy}[1]{\textsf{\mdseries #1}\xspace}
\newcommand{\csfs}{\casestudy{forestall}}
\newcommand{\csiot}{\casestudy{iot-dev}}
\newcommand{\csadmin}{\casestudy{gain-admin}}

\newcommand{\imitator}{\MakeUppercase{imitator}}
\newcommand{\uppaal}{\textsc{Uppaal}\xspace}

\Crefname{section}{Sec.}{Secs.}
\crefname{section}{section}{sections}
\Crefname{subsection}{Sec.}{Secs.}
\crefname{subsection}{section}{sections}
\Crefname{equation}{Eq.}{Eqs.}
\crefname{equation}{equation}{equations}
\Crefname{definition}{Def.}{Defs.}
\crefname{definition}{definition}{definitions}
\Crefname{algorithm}{Alg.}{Algs.}
\crefname{algorithm}{algorithm}{algorithms}
\crefname{table}{table}{tables}
\Crefname{figure}{Fig.}{Figs.}
\crefname{figure}{figure}{figures}

\makeatletter
\renewcommand{\paragraph}{\@startsection{paragraph}{4}{0pt}%
  {.8ex plus 0.2ex minus 0.2ex}%
  {-0.5em}%
  {\bfseries}}
\makeatother

\newcommand{\marginX}{\marginnote{\huge{\quad\textbf{!}\quad}}}

\title{Hackers vs.\ Security: Attack-Defence Trees\\
	as Asynchronous Multi-Agent Systems\thanks{This work was partially funded by the NWO project SEQUOIA (grant~15474), EU
project SUCCESS (102112) and the PHC van Gogh PAMPAS. The work of Arias and Petrucci
has been supported
by the BQR project AMoJAS.}}
\author{Jaime~Arias\inst{1}\and Carlos~E.~Budde\inst{4}\and
Wojciech~Penczek\inst{2,3}\and Laure~Petrucci\inst{1}\and
Mari\"{e}lle~Stoelinga\inst{4,5}
}
\institute{
	LIPN, CNRS UMR 7030, Universit\'{e} Paris 13, Sorbonne Paris Cit\'{e},\\
	Villetaneuse, France \and
	Institute of Computer Science, PAS, Warsaw, Poland \and
	University of Natural Sciences and Humanities, II, Siedlce, Poland \and
	Formal Methods and Tools, University of Twente, Enschede, The Netherlands \and
	Department of Software Science, Radboud University, Nijmegen, The Netherlands
}
\begin{document}
\maketitle

\begin{abstract}
\lp{We can think of a better title}
\ceb{Updated by Mariëlle}
Attack-Defence Trees (\ADT/s) are well-suited to assess possible attacks to
systems and the efficiency of counter-measures. In this paper, we first enrich
the available constructs with reactive patterns that cover further security
scenarios, and equip all constructs with attributes such as time and cost to
allow quantitative analyses. Then, \ADT/s are modelled as (an extension of)
Asynchronous Multi-Agents Systems---\EAMAS/. The \ADT/--\EAMAS/ transformation
is performed in a systematic manner that ensures correctness.  The
transformation allows us to quantify the impact of different agents
configurations on metrics such as attack time. Using \EAMAS/ also permits
parametric verification: we derive constraints for property satisfaction. Our
approach is exercised on several case studies using the \uppaal\ and \imitator\
tools.
\end{abstract}

\section{Introduction}
\label{sec:intro}

Over the past ten years of security analysis, a multitude of formalisms has been
developed to study interactions between attacker and defender parties.
Attack-defence trees stand out among these, as a graphical, straightforward
formalism of great modelling versatility.
However, research is thus far focused on bipartite graph characterisations,
where nodes belong to either the attacker or defender party. In real security
scenarios each of these parties may be formed of multiple agents. Agents
distribution over the tree nodes, \ie which agent performs which task for which
goal, determines not only the \emph{performance} but also the \emph{feasibility}
of an attack or defence strategy. Studying these distributions is thus
paramount.

To analyse the impact of different agents distributions, we model attack-defence
trees in an agent-aware formalism. Our approach permits quantifying performance
metrics (\eg cost and time) of attack/defence strategies under distinct agents
coalitions.  Employing modern verification tools such as \imitator, we determine
the feasibility of coalition strategies, and synthesise the value of the
attributes that make them feasible, such as the maximum time allowed for a
defence mechanism to be triggered. This way, we make an important step towards
richer \ADT/ representations for real-world security analysis.

\paragraph{Contributions.}
Concretely, in this paper we introduce:
\begin{enumerate*}[($i$)]
	\item a unified scheme for \ADT/ representation with counter- and sequential
	      operators, including \emph{a new construct};
	\item a formalism---\EAMAS/---to model \ADT/s where \emph{all nodes} have
	      \emph{attributes}, and can be operated by different \emph{agents};
	\item compositional and sound \emph{transformation rules} from \ADT/ to \EAMAS/;
	\item measurements of the impact of different \emph{agents coalitions} on
	      attack performance metrics, such as time, exercised on several case studies;
	\item \emph{synthesis of \ADT/ attributes}---by encoding them as \EAMAS/
	      parameters---for feasibility analysis of attack/defence strategies.
\end{enumerate*}

\paragraph{Outline.}
The paper is structured as follows.
In \Cref{sec:ADT} we review the basic notions of \ADT/s, using a unified
syntactic and semantic representation. \Cref{sec:AMAS} reviews the \AMAS/
formalism, which we extend in \Cref{sec:EAMAS} to model \ADT/s enriched with
attributes.  \Cref{sec:patterns} presents graph-based transformation patterns to
model each \ADT/ construct as an (extended) \AMAS/ model. Such compositional
modelling avoids state space issues; we prove it correct for our semantics and
illustrate it in our running example. The effectiveness of our approach is
demonstrated in \Cref{sec:expe}, by using two state-of-the-art verification
tools to quantitatively analyse three case studies from the literature.  This
work concludes in \Cref{sec:conclu}, where we finally draw possible lines of
future research.

\section{Attack-Defence Trees}
\label{sec:ADT}


\subsection{The basic \ADT/ model and related work}
\label{sec:adt:basic}

\emph{Attack trees} are well-known tree-like representations of attack scenarios,
that allow for evaluating the security of complex systems to the desired degree of
refinement \cite{MO05}. The root of the tree is the attacker's goal, and
the children of a node represent refinements of the node's goal into sub-goals.
The tree leaves are (possibly quantified) foreseeable attacker's actions.  For
instance, if the attack goal is to \uline{S}teal \uline{J}ewels from a museum
(\texttt{SJ}), where burglars must \uline{b}reak \uline{i}n (attack
leaf~\texttt{bi}) and \uline{f}orce a \uline{d}isplay (\texttt{fd}), the attack
tree is: $\mathtt{SJ}=\gateAND(\mathtt{bi},\mathtt{fd})$.

\begin{wrapfigure}[9]{r}[0pt]{.24\linewidth}
	\centering
	\vspace{-4.8ex}
	\scalebox{.75}{\begin{tikzpicture}
		[every node/.style={ultra thick,draw=red,minimum size=6mm}]
		\node[and gate US,point up,logic gate inputs=ni] (SJS)
		{\rotatebox{-90}{\texttt{SJS}}};
		\node[rectangle,draw=Green,minimum size=8mm, below = 4mm of SJS.west, xshift=10mm] (p)
		{\texttt{p}};
		\draw (p.north) -- ([yshift=1.8mm]p.north) -| (SJS.input 2);
		\node[and gate US,point up,logic gate inputs=nn, below = 8mm of SJS.west, yshift=14mm] (SJ)
		{\rotatebox{-90}{\texttt{SJ}}};
		\draw (SJ.east) -- ([yshift=0.6mm]SJ.east) -| (SJS.input 1);
		\node[state, below = 3.5mm of p.south] (fd)
		{\texttt{fd}};
		\draw (fd.north) -- ([yshift=1.6mm]fd.north) -| (SJ.input 2);
		\node[state, left = 1.08cm of fd.west, yshift=-.2mm] (bi)
		{\texttt{bi}};
		\draw (bi.north) -| ([xshift=1mm]SJ.input 1);
	\end{tikzpicture}}
	\caption{\small Steal Jewels\label{fig:adt:stjews}}
\end{wrapfigure}
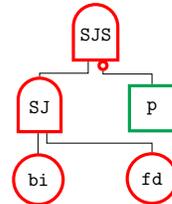

\emph{Attack-defence trees} (\ADT/s, \cite{KordyMRS10}) are an extension of
attack trees including possible counteractions of a defender, thus representing
security scenarios as an interplay between attacker's and defender's actions.
This can model mechanisms triggered by the occurrence of specified opposite
actions. So for instance, the jewels burglary will succeed
(\texttt{SJS}) only if all attack actions are performed, and the alerted
\uline{p}olice (\texttt{p}) doesn't stop them:
\mbox{$\mathtt{SJS = \gateNAND(SJ,p)}$}, see \Cref{fig:adt:stjews}.

Countering actions emerged recently to express pointwise
interactions between opposite players. This can be reactive, such as the police
above, or passive, such as message encryption to avoid
eavesdropping in open channels. The intuition behind the counter actions in
\cite{KordyMRS10} can be described as follows:
\begin{itemize}[label=$\blacktriangleright$,topsep=.3ex,itemsep=.3ex,leftmargin=*]
	\item \emph{counter defence} $\ActAttack=\gateNAND(\actAttack,\actDefence)$:
		  if attack $\actAttack$ succeeds and the countering defence
	      $\actDefence$ fails, then the attack goal~$\ActAttack$ is successful;
	\item \emph{counter attack} $\ActDefence=\gateNAND(\actDefence,\actAttack)$:
	      if defence $\actDefence$ happens and attack $\actAttack$ fails,
	      then the defence goal~$\ActDefence$ is successful.
\end{itemize}

Formally, we give Boolean semantics to the nodes composing \ADT/s as shown in
\Cref{tab:adt}. As with counter defence/attack above, constructs are symmetric
for attack or defence goals; thus the \namecref{tab:adt} shows only a selection.
In the graphical 
representations, triangular nodes indicate (arbitrary) subtrees can be the
children of a gate, whereas circular
(resp.\ rectangular) nodes represent attack (resp.\ defence) leaves. In this
way, with \Cref{tab:adt} we extend \cite{KordyMRS10} with \gateAND/\gateOR gates
along the lines of~\cite{attack_FASE18}.  We further extend notation---in
accordance with these semantics---for nodes that represent countering a defence
or an attack. Thus, for instance, the \ADT/ jewels burglary example has
the semantics $\mathtt{SJS = bi \land fd \land \lnot p}$, graphically
depicted as in \Cref{fig:adt:stjews}.

In \Cref{tab:adt} we also introduce operators for a \emph{choice} between a
successful attack and a failing defence (named \emph{no defence}), and
vice-versa (\emph{inhibiting attack}). These constructs are less usual in the
literature \cite{KPCS13}, yet they model realistic scenarios such as attack
goals succeeding by security negligence rather than by performing an elaborated
(and costly!) attack. This is of interest for quantitative analyses of \eg
cost and probability, and thus we include them in \Cref{tab:adt}.

\LongVersion{Notice that attack-defence formalisms that include the negation
operator can express no defence and inhibiting attack: in that sense
\cite{Zaruhi_pareto_2015} encode these via their ``changing player operator''
($\sim$), where no defence is $\sim\!(\mathrm{counter~attack})$ and
inhibiting attack is $\sim\!(\mathrm{counter~defence})$. \ja{I'm not sure
about the equivalence.}}\ja{Can we compare sequential gates with the
well-know term "short-circuit evaluation"?}


\LongVersion{
Moreover, we consider sequential order operators, for which there is no
standard behavioural interpretation in attack-defence trees.  In the survey of
\cite{KPCS13} several ways to model order-dependent events are reported,
\eg the sequential enforcement (\gate{seq}) of dynamic fault trees
\cite{Kha09,JGKS16}, time dependent ordered-\gateAND in enhanced attack trees
\cite{CY07}, priority- or sequential-\gateAND, etc.  Jhawar et al.\ give a
comprehensive overview of different interpretations for sequentially ordered
conjunction in the context of attack trees in
\cite{JKMR15}. They propose a ``series-parallel graph semantics,'' with whose
resulting behaviour we agree, for which a sequential conjunction attack
(\gateSAND) is successful as soon as all children have occurred in the required
order.

This semantics abstracts away the exact time of occurrence of events, that is,
transition attributes (such as the probability or the cost of an attack) are
not a function of time. With few exceptions (\eg \cite{CY07,FHRM14}) this is
the preferred approach in the literature
\cite{SSSW98,SHJLW02,BLPSW06,Kha09,KPCS13,JKMR15,attack_FASE18}.  Notice that
\gateSAND{s} in \cite{JKMR15} describe the order in which actions will
\emph{certainly} take place rather than a potential outcome order of events.
Their role is to assert the sequentiality of events and rule out parallel
executions, in a sense similarly to \gate{seq} in dynamic fault trees and in
opposition to priority-\gateAND \cite{Kha09}.
Thanks to this time-abstract interpretation of the outcome of a \gateSAND, we
are able to provide Boolean semantics to it in \Cref{tab:adt}.
}

\ShortVersion{
Moreover, we consider sequential order operators, which lack a standard
interpretation for \ADT/s. 
The survey of \cite{KPCS13} lists several ways to model order-dependent
events: sequential enforcement, time dependent ordered-\gateAND, priority- or
sequential-\gateAND, etc. For attack trees, in the sequentially ordered
conjunction (\gateSAND) of \cite{JKMR15}, attacks
succeed when all children occurred in the required order.  This abstracts
away the exact time of occurrence of events; with few exceptions,
\eg~\cite{CY07}, this is the preferred approach in the literature
\cite{SSSW98,BLPSW06,KPCS13,JKMR15,attack_FASE18}.
Therefore \gateSAND{s} describe the order in which actions
\emph{will} take place, rather than a potential order of events. Their role is
to enforce \emph{sequential events} and rule out parallel executions, which is a
fundamental construct when working with multi-agent systems.  For instance,
Steal Jewels (\texttt{SJ}) in \Cref{fig:adt:stjews} is modelled with an \gateAND
gate. Say that breaking in (\texttt{bi}) takes 10~min and forcing the display
(\texttt{fd}) 5~min. If two attackers operate in parallel, analysing the \ADT/
can conclude that \texttt{SJ} succeeds after a 10~min attack. But \texttt{fd}
depends logically on \texttt{bi}, since the display can only be forced after
breaking in. Using instead a \gateSAND gate for \texttt{SJ} enforces this
sequentiality so that attacks cannot take less than 15~min. We integrate such
\gateSAND gates in our \ADT/ framework, as the \emph{sequential and
attack/defence} in \Cref{tab:adt}.
}


We further introduce \gateSNAND gates: sequential gates that have attacks
and defences as children.  To the best of our knowledge, this is novel in a
typed setting where subtrees (rather than leaves) can be assigned attack/defence
goals.  This contribution is conservative: it extends the \gateSAND gates of
\cite{JKMR15} to coincide with previous works on sequential operators in
defence-aware representations, \eg attack-defence diagrams \cite{HermannsKKS16}.
We distinguish two scenarios: a successful attack followed by a failed defence
(named \emph{failed reactive defence} in \Cref{tab:adt}), and vice versa.  We
disregard the second scenario as uninteresting---it models defence goals which
depend on attacks failing by themselves---and focus on the first one.
\gateSNAND{s} then model an attack goal that must overcome some reactive
defence, triggered only after the incoming attack has been detected.

\LongVersion{
\begingroup
\def\light{\textsl{light}\xspace}
\def\heavy{\textsl{heavy}\xspace}
By means of an example consider a server which can operate using one of two data
encryption algorithms, \light or \heavy, where \light is a fast stream cipher
of moderate complexity and \heavy is a super-secure but also slow block cipher.
The company offering the hosting service is reluctant to use \heavy encryption
since the high data access latency will favour the competition; unless
malicious activity is detected by their online firewalls, which will switch the
system to emergency mode and swap \light for \heavy encryption during the next
eight hours---time enough for the security team to show in scene and identify
the threat.
An attacker trying to access private data in the server will use a script that
is effective against \light encryption only: the outcome of the attack will
depend on the subsequent timely activation of the emergency mode that switches
to \heavy encryption.  In particular, if the company decided to use \heavy
encryption (regardless of the attack) then the attack will fail, but this will
not be known by the attacker \emph{until he has launched the attack} and it
was---unsuccessfully---completed.
\endgroup
}

\ceb{Which structure can the children of \gateSNAND have? \textbf{No shared defences} because we lose compositionality (say, this \gateSNAND can't know that the other \gateSNAND requires the shared defence to occur later) Can we have general attack (on the left) and defence (on the right) refinements? Or are there further restrictions to those subtrees?}

\ifdefined\VersionLong
\input{tables/tabADT}
\else
\vspace{-2ex}

\begin{table}[h!]
\centering
\scalebox{0.7}{
\large
\def\nodeattack{\node[state,draw=red,ultra thick]}
\def\nodedefence{\node[rectangle,draw=Green,ultra thick]}
\def\nodesubtree{\node[isosceles triangle, isosceles triangle apex angle=70, shape border rotate=90, minimum size=6.5mm]}
%
\begin{tabular}{cccc}
{\bf Name} & {\bf Domains} & {\bf Graphics} & {\bf Semantics}
\\[.5ex]\hline

\vspace{.5ex}
Attack & $\actAttack\in\actAttackSet$
	& \parbox[c]{2cm}{\centering
	\vspace{1mm}
	\begin{tikzpicture}
		\normalsize
		\nodeattack[minimum size=6mm] (LA) {$\actAttack{}$};
	\end{tikzpicture}
	\vspace{1mm}
	}
	& $\actAttack$
\\
Defence & $\actDefence\in\actDefenceSet$
	& \parbox[c]{2cm}{\centering
	\vspace{1mm}
	\begin{tikzpicture}
		\normalsize
		\nodedefence[minimum size=6mm] (LD) {$\actDefence{}$};
	\end{tikzpicture}
	\vspace{1mm}
	}
	& $\actDefence$
\\
And attack
	& $\ActAttack,\actAttack{}_1,\cdots,\actAttack{}_n\in\actAttackSet{}$
	& \parbox[c]{4cm}{\centering
	\vspace{1mm}

\begingroup
\def\nodesubtree{\node[isosceles triangle, isosceles triangle apex angle=70, shape border rotate=90, minimum size=6.5mm]}

\begin{tikzpicture}
	[every node/.style={ultra thick,draw=red,minimum size=6mm}]
	\normalsize
	\node[and gate US,point up,logic gate inputs=nn] (A)
		{\rotatebox{-90}{\large$\ActAttack$}};
	\nodesubtree at (-1,-1.3) (a1) {$\!\actAttack_1\!\!$};
	\nodesubtree at ( 1,-1.3) (an) {$\!\actAttack_n\!\!$};
	\node[draw=none] at (0,-1.3) {$\cdots$};
	\draw (a1.north) -- ([yshift=1.5mm]a1.north) -| (A.input 1);
	\draw (an.north) -- ([yshift=1.3mm]an.north) -| (A.input 2);
\end{tikzpicture}
\endgroup\!\!
	\vspace{1mm}
	}
	& $\ActAttack=\actAttack{}_1\land\cdots\land\actAttack{}_n$
\\
Or defence
	& $\ActDefence{},\actDefence{}_1,\cdots,\actDefence{}_n\in\actDefenceSet{}$
	& \parbox[c]{4cm}{\centering
	\vspace{1mm}

\begingroup
\def\nodesubtree{\node[isosceles triangle, isosceles triangle apex angle=70, shape border rotate=90, minimum size=6.5mm]}

\begin{tikzpicture}
	[every node/.style={ultra thick,draw=Green,minimum size=6mm},every state/.style={rectangle}]
	\normalsize
	\node[or gate US,point up,logic gate inputs=nn] (D)
		{\rotatebox{-90}{\large$\ActDefence$}};
	\nodesubtree at (-1,-1.3) (d1) {$\!\actDefence_1\!\!$};
	\nodesubtree at ( 1,-1.3) (dn) {$\!\actDefence_n\!\!$};
	\node[draw=none] at (0,-1.3) {$\cdots$};
	\draw (d1.north) -- ([yshift=1.5mm]d1.north) -| (D.input 1);
	\draw (dn.north) -- ([yshift=1.3mm]dn.north) -| (D.input 2);
\end{tikzpicture}
\endgroup\!\!
	\vspace{1mm}
	}
	& $\ActDefence{}=\actDefence{}_1\lor\cdots\lor\actDefence{}_n$
\\
Counter defence
	& \parbox[c]{.2\linewidth}{%
		$\begin{aligned}
		\ActAttack{},\actAttack{} &\in \actAttackSet{}\\
		\actDefence{}             &\in \actDefenceSet{}
		\end{aligned}$
	}
	& \parbox[c]{4cm}{\centering
	\vspace{1mm}

\begingroup
\def\nodesubtree{\node[isosceles triangle, isosceles triangle apex angle=70, shape border rotate=90, minimum size=6.5mm]}

\begin{tikzpicture}
	[every node/.style={ultra thick,draw=red,minimum size=6mm}]
	\normalsize
	\node[and gate US,point up,logic gate inputs=ni] (A)
		{\rotatebox{-90}{\large$\ActAttack$}};
	\nodesubtree             at (-1,-1.36) (a) {$\actAttack$};
	\nodesubtree[draw=Green] at ( 1,-1.32) (d) {$\!\actDefence\!$};
	\draw (a.north) -- ([yshift=1.4mm]a.north) -| (A.input 1);
	\draw (d.north) -- ([yshift=1.5mm]d.north) -| (A.input 2);
\end{tikzpicture}
\endgroup\!\!
	\vspace{1mm}
	}
	& $\ActAttack=\actAttack{}\land\neg\actDefence{}$
\\
No defence
	& \parbox[c]{.2\linewidth}{%
		$\begin{aligned}
		\ActAttack{},\actAttack{} &\in \actAttackSet{}\\
		\actDefence{}             &\in \actDefenceSet{}
		\end{aligned}$}
	& \parbox[c]{4cm}{\centering
	\vspace{1mm}

\begingroup
\def\nodesubtree{\node[isosceles triangle, isosceles triangle apex angle=70, shape border rotate=90, minimum size=6.5mm]}

\begin{tikzpicture}
	[every node/.style={ultra thick,draw=red,minimum size=6mm}]
	\normalsize
	\node[or gate US,point up,logic gate inputs=ni] (A)
		{\rotatebox{-90}{\large$\ActAttack$}};
	\nodesubtree             at (-1,-1.28) (a) {$\actAttack$};
	\nodesubtree[draw=Green] at ( 1,-1.24) (d) {$\!\actDefence\!$};
	\draw (a.north) -- ([yshift=1.4mm]a.north) -| (A.input 1);
	\draw (d.north) -- ([yshift=1.5mm]d.north) -| (A.input 2);
\end{tikzpicture}
\endgroup\!\!
	\vspace{1mm}
	}
	& $\ActAttack=\actAttack{}\lor\neg\actDefence{}$
\\
Inhibiting attack
	& \parbox[c]{.2\linewidth}{%
		$\begin{aligned}
		\ActDefence{},\actDefence{} &\in \actDefenceSet{}\\
		\actAttack{}                &\in \actAttackSet{}
		\end{aligned}$}
	& \parbox[c]{4cm}{\centering
	\vspace{1mm}

\begingroup
\def\nodesubtree{\node[isosceles triangle, isosceles triangle apex angle=70, shape border rotate=90, minimum size=6.5mm]}

\begin{tikzpicture}
	[every node/.style={ultra thick,draw=Green,minimum size=6mm}]
	\normalsize
	\node[or gate US,point up,logic gate inputs=ni] (D)
		{\rotatebox{-90}{\large$\ActDefence$}};
	\nodesubtree           at (-1,-1.24) (d) {$\!\actDefence\!$};
	\nodesubtree[draw=red] at ( 1,-1.28) (a) {$\actAttack$};
	\draw (d.north) -- ([yshift=1.5mm]d.north) -| (D.input 1);
	\draw (a.north) -- ([yshift=1.4mm]a.north) -| (D.input 2);
\end{tikzpicture}
\endgroup\!\!
	\vspace{1mm}
	}
	& $\ActDefence=\actDefence{}\lor\neg\actAttack{}$
\\
\parbox[c]{.3\linewidth}{\centering Sequential\\and attack}
	& $\ActAttack,\actAttack{}_1,\cdots,\actAttack{}_n\in\actAttackSet{}$
	& \parbox[c]{4cm}{\centering
	\vspace{1mm}

\begingroup
\def\nodesubtree{\node[isosceles triangle, isosceles triangle apex angle=70, shape border rotate=90, minimum size=6.5mm]}

\begin{tikzpicture}
    [every node/.style={ultra thick,draw=red,minimum size=6mm}]
	\normalsize
	\node[and gate US,point up,logic gate inputs=nn, seq=4pt] (A)
		{\rotatebox{-90}{\large$\ActAttack$}};
	\nodesubtree at (-1,-1.3) (a1) {$\!\actAttack_1\!\!$};
	\nodesubtree at ( 1,-1.3) (an) {$\!\actAttack_n\!\!$};
	\node[draw=none] at (0,-1.3) {$\cdots$};
	\draw (a1.north) -- ([yshift=1.5mm]a1.north) -| (A.input 1);
	\draw (an.north) -- ([yshift=1.5mm]an.north) -| (A.input 2);
\end{tikzpicture}
\endgroup\!\!
	\vspace{1mm}}
	& $\ActAttack=\actAttack{}_1\land\cdots\land\actAttack{}_n$
\\
\parbox[c]{.3\linewidth}{\centering Failed reactive\\defence}
	& \parbox[c]{.2\linewidth}{%
		$\begin{aligned}
		\ActAttack{},\actAttack{} &\in \actAttackSet{}\\
		\actDefence{}             &\in \actDefenceSet{}
		\end{aligned}$}
	& \parbox[c]{4cm}{\centering
	\vspace{1mm}

\begingroup
\def\nodesubtree{\node[isosceles triangle, isosceles triangle apex angle=70, shape border rotate=90, minimum size=6.5mm]}

\begin{tikzpicture}
	[every node/.style={ultra thick,draw=red,minimum size=6mm}]
	\normalsize
	\node[and gate US,point up,logic gate inputs=ni, seq=9pt] (A)
		{\rotatebox{-90}{\large$\ActAttack$}};
	\nodesubtree             at (-1,-1.36) (a) {$\actAttack$};
	\nodesubtree[draw=Green] at ( 1,-1.32) (d) {$\!\actDefence\!$};
	\draw (a.north) -- ([yshift=1.4mm]a.north) -| (A.input 1);
	\draw (d.north) -- ([yshift=1.5mm]d.north) -| (A.input 2);
\end{tikzpicture}
\endgroup\!\!
	\vspace{1mm}
	}
	& $A = a \land \lnot d$
\\
\end{tabular}
}
\caption{\ADT/ constructs (selection) and their semantics\label{tab:adt}}
\end{table}

\vspace{-5ex}
\fi

\subsection{Attributes and agents for \ADT/s}
\label{sec:adt:attributes}

Attributes (also ``parameters'' and ``gains''
\cite{BLPSW06,Zaruhi_pareto_2015,attack_FASE18}) are numeric properties of
attack/defence nodes that allow for quantitative analyses.  Typical attributes
include cost, time, and probability: in the Steal Jewels example, the 10~min it
takes to break in is a time attribute of this attack leaf. If all nodes in an
\ADT/ have a time attribute, one can compute the min/max time required by the
attack goal, whereas in general attributes are associated to tree leaves.
As in \cite{ALRS19}, attributes can be parameters over which we synthesize constraints
indicating which attribute values can lead to a successful attack. 

\paragraph*{General attributes.}
We extend attributes to all constructs (\ie nodes) in \ADT/s, because a
construct may not be fully described by just its children. An attribute is then
given by a node's \emph{initial} (aka \emph{intrinsic}) \emph{value}, and a
\emph{computation function}.
For example, refine \texttt{bi} to be an \gateAND gate with children
\uline{p}ick \uline{m}ain \uline{l}ock (\texttt{pml}, 7~min) and \uline{s}aw
\uline{p}adlock (\texttt{sp}, 2~min). Then it may still take an extra minute to
enter stealthily and locate the correct display, to be counted after \texttt{pml}
and \texttt{sp} have succeeded.  In general, when the goal of a gate is
successful, its attribute value is the result of its computation function
applied to its intrinsic value and to the attributes of its children. For
\texttt{bi}, if two burglars cooperate, the computation function is
$\mathit{init\_time}(\mathtt{bi})
 + \max(\mathit{init\_time}(\mathtt{pml}),\mathit{init\_time}(\mathtt{sp}))$.
This allows for flexibility in describing different kinds of attributes, and
gains special relevance when considering coalitions of agents, as we will
further illustrate in \Cref{sec:adt:example}.

\paragraph*{Agents.}
Each action described by an \ADT/ construct can be performed by a
particular \emph{agent}. Different attacks/defences could be handled by one or
multiple agents, which allows to express properties on agents coalitions. For
instance, in the original Steal Jewels example of \Cref{fig:adt:stjews}, the
minimal number of burglars
required to minimise the \texttt{SJS} attack time is two:
one to do \texttt{bi} and another to parallelly perform \texttt{fd}. If
the \texttt{SJ} gate is changed to a \gateSAND, then one burglar suffices, since
\texttt{bi} and \texttt{fd} cannot be parallelised. Upon using the refinement
$\mathtt{bi = \gateAND(pml,sp)}$, then again a coalition of two burglars
minimises the attack time, since \texttt{pml} and \texttt{sp} can be
parallelised.
Each node in the \ADT/ will thus be assigned to an agent, and a single agent can
handle multiple nodes. In the general case, the only constraint is that no agent
handles both attack and defence nodes. As shown above, specific \ADT/s might
enforce additional constraints.

\paragraph*{Conditional counter measures.}
It may happen that a countering node has a successful or unsuccessful outcome
depending on the attributes of its children. We therefore associate
\emph{conditions} with countering nodes, which are Boolean functions over the
attributes of the \ADT/. When present, the condition then comes as an additional
constraint for the node operation to be successful.

\subsection{Example: Treasure hunters}
\label{sec:adt:example}

\begin{wrapfigure}[24]{r}[0pt]{.4\linewidth}
	\vspace{-3ex}
	\caption{The treasure hunters\label{fig:adt:treasure}}
	\vspace{-1ex}
	{  
	  \qquad%
	  \scalebox{.7}{
	  \begin{tikzpicture}
		[every node/.style={ultra thick,draw=red,minimum size=6mm}]
		\node[and gate US,point up,logic gate inputs=ni] (ca)
		{\rotatebox{-90}{\texttt{TS}}};
		\node[rectangle,draw=Green,minimum size=8mm, below = 5mm of ca.west, xshift=10mm] (d) {\texttt{p}};
		\draw (d.north) -- ([yshift=0.28cm]d.north) -| (ca.input 2);
		\node[and gate US,point up,logic gate inputs=nn, seq=4pt, below = 9mm of ca.west, yshift=14mm] (A)
		{\rotatebox{-90}{\texttt{TF}}};
		\draw (A.east) -- ([yshift=0.15cm]A.east) -| (ca.input 1);
		\node[and gate US,point up,logic gate inputs=nn, below = 9mm of A.west, yshift=14mm] (a1)
		{\rotatebox{-90}{\texttt{ST}}};
		\draw (a1.east) -- ([yshift=0.15cm]a1.east) -| (A.input 1);
		\node[state, below = 4mm of a1.west, xshift=-5mm] (a1n) {\texttt{b}};
		\draw (a1n.north) -- ([yshift=0.15cm]a1n.north) -| (a1.input 1);
		\node[state, below=4mm of a1.west, xshift=5mm] (a11) {\texttt{f}};
		\draw (a11.north) -- ([yshift=0.15cm]a11.north) -| (a1.input 2);
		\node[or gate US,point up,logic gate inputs=nn, below = 10mm of A.west, yshift=-7mm] (a2)
		{\rotatebox{-90}{\texttt{GA}}};
		\draw (a2.east) -- ([yshift=0.15cm]a2.east) -| (A.input 2);
		\node[state, below = 4mm of a2.west, xshift=-5mm] (a21) {\texttt{h}};
		\draw (a21.north) -- ([yshift=0.15cm]a21.north) -| (a2.input 1);
		\node[state, below=4mm of a2.west, xshift=5mm] (a2n) {\texttt{e}};
		\draw (a2n.north) -- ([yshift=0.15cm]a2n.north) -| (a2.input 2);
	  \end{tikzpicture}
	  }
	  \subcaption{\ADT/}
	  \label{fig:ths:adt}
	}
	~\\
	{  
	  \scalebox{.8}{\parbox{\linewidth}{%
	    ~%
		\begin{tabular}{r@{~}l@{~\;}r@{~\;}r}
			\multicolumn{2}{l}{\bf Name} & {\bf Cost} & {\bf Time} \\
			\hline
			\texttt{TS} & (treasure stolen)  & &\\
			\texttt{p}  & (police)           & 100\,\euro{} & 10 min\\
			\texttt{TF} & (thieves fleeing)  & & \\
			\texttt{ST} & (steal treasure)   &              & 2 min \\
			\texttt{b}  & (bribe gatekeeper) & 500\,\euro{} & 1 h \\
			\texttt{f}  & (force arm. door)  & 100\,\euro{} & 2 h \\
			\texttt{GA} & (go away)          & & \\
			\texttt{h}  & (helicopter)       & 500\,\euro{} & 3 min \\
			\texttt{e}  & (emergency exit)   &              & 10 min
		\end{tabular}
		~\\[.2ex]
		\uline{\textbf{Condition for \texttt{TS}}}:\\[.2ex]
		\mbox{$\initTime{\mathtt{p}} > \Time{\mathtt{ST}} + \Time{\mathtt{GA}}$}
		\vspace{-.7ex}
	  }}
	  \subcaption{Attributes of nodes}
	  \label{fig:ths:attributes}
	}
\end{wrapfigure}
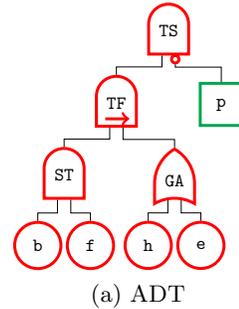

We present a simple running example in \Cref{fig:adt:treasure} that extends
\Cref{fig:adt:stjews}, featuring thieves that try to steal a treasure in a
museum. To achieve their goal, they first must access the treasure room, which
involves distracting and \uline{b}ribing a guard ({\tt b}), and \uline{f}orcing
the secure door ({\tt f}).
%
%
Both actions are costly and take some time. Two coalitions are possible:
it may be that a single thief has to carry out both actions, or there could
be two thieves that collaborate to parallelise \texttt{b} and \texttt{f}.

After these actions succeed the thieves can \uline{s}teal the \uline{t}reasure
(\gate{ST}), which takes a little time for opening its display stand and putting
it in a bag. Then
%
the \uline{t}hieves are ready to \uline{f}lee (\gate{TF}), but
to do that they have to choose an escape
route to \uline{g}o \uline{a}way
(\gate{GA}): this can be a spectacular escape in a \uline{h}elicopter ({\tt h}),
or a mundane escape via the \uline{e}mergency exit ({\tt e}). The helicopter is
expensive but fast while the fire exit is slower but at no cost. These values
will allow for experimenting different solutions.  Furthermore, the time to
perform a successful escape can be inversely proportional to the number of
agents involved in the robbery. This is easy to encode via a computation
function in the gate \gate{GA}.

As soon as the treasure room is penetrated (\ie after \texttt{b} and \texttt{f}
but before \gate{ST}) an alarm goes off at the police station, so while the
thieves flee the \uline{p}olice hurries to intervene ({\tt p}).  The treasure is
then successfully stolen iff the thieves have fled and the police failed to
arrive (\eg because
they were on strike%
\footnote{\url{https://en.wikipedia.org/wiki/2013_police_revolts_in_Argentina}})
or does so too late. This last possibility is captured by the additional
condition associated with the \uline{t}reasure \uline{s}tolen gate ({\tt TS}),
which states that the arrival time of the police must be greater than the time
for the thieves to steal the treasure and go away.

\section{Asynchronous Multi-Agent Systems}
\label{sec:AMAS}

\def\A{A}
\def\PV{\mathit{PV}}
\def\States{S}
\def\state{s}
\def\trans#1{\stackrel{#1}{\longrightarrow}}
\def\Agent{\ensuremath{\mathit{Agent}}}

\emph{Asynchronous Multi-Agent Systems} (\AMAS/, \cite{AAMASWJWPPDAM2018a}) are
a modern semantic model for the study of agents' strategies in asynchronous
systems. They offer a (succinctly represented) analysis framework capable of
efficient reachability checks even on non-trivial models
\cite{AAMASWJWPPDAM2018a}. By linking protocols to agents, \AMAS/ are a natural
compositional formalism for analysing multi-agent systems.

Technically, \AMAS/ are similar to networks of automata that synchronize on
shared actions, and interleave local transitions to execute asynchronously
\cite{Fagin95knowledge,LomuscioPQ10a,AAMASWJWPPDAM2018a}. Works that
give semantics to attack-trees via automata usually do not consider defence nodes
\cite{CY07,attack_FASE18}. The attack-defence diagrams of \cite{HermannsKKS16}
use stochastic timed automata as a semantic base: this grants them ample
expressive power, but also entails large state spaces and undecidable
reachability in the general case%
\LongVersion{.  This must be circumvented resorting to approximate results
\cite{HHH14} or simulation based approaches \cite{HermannsKKS16}. Instead,
reachability properties in \AMAS/ models are decidable, and its succinct state
space representation allows to easily check reachability even on non-trivial
models.}%
\ShortVersion{, which \AMAS/ circumvents.}

\begin{definition}[Asynchronous Multi-Agent Systems, \cite{AAMASWJWPPDAM2018a}]
\label{def:amas}
An \emph{asyn\-chronous multi-agent system (\AMAS/)} consists of $n$
\emph{agents} $A = \{1,\dots,n\}$, where each agent has an associated tuple
$A_i =(L_i, \iota_i, Act_i, P_i, T_i)$ including
\begin{enumerate*}[(i)]
\item a set of \emph{local states} $L_i=\{l_i^1, l_i^2,\dots,l_i^{n_i}\}$;
\item an \emph{initial state} $\iota_i\in L_i$;
\item a set of \emph{actions} $Act_i=\{a_i^1,a_i^2,\ldots, a_i^{m_i}\}$;
\item a \emph{local protocol} $P_i \colon L_i \to 2^{Act_i}$ which selects the actions
available at each local state; and
\item  a (partial) \emph{local transition function}
$T_i \subseteq L_i \times Act_i \times L_i$  \st
$(l_i,a,l'_i) \in T_i$ for some $l'_i \in L_i$ iff $a \in P_i(l_i)$.
\end{enumerate*}
\end{definition}

Sets $Act_i$ need not be disjoint.  $Act = \bigcup_{i \in \A} Act_i$ and $Loc =
\bigcup_{i \in \A} L_i$ are the set of all actions and all local states, respectively.
For each action $a \in Act$, the set $\Agent(a) = \left\{ i\in\A \mid a \in
Act_i\right \}$ contains all agents that can perform action $a$.
The parallel composition of \AMAS/ is given by Interleaved Interpreted
Systems, which extend \AMAS/ with variables and define \emph{global states} and
\emph{global transitions}.
\ceb{This is defined in \cite{AAMASWJWPPDAM2018a} and, unlike \Cref{def:amas},
doesn't seem crucial for defining \EAMAS/ (the ``theoretical contribution'' of this work) in \Cref{sec:EAMAS}.
Consider removal if we're short on space.}

\begin{definition}[Interleaved Interpreted System, \cite{AAMASWJWPPDAM2018a}]
Let $\PV$ be a set of propositional variables.
An \emph{interleaved interpreted system (IIS)}%
\LongVersion{, or a \emph{model},}
is an \AMAS/ extended with%
\LongVersion{ the following elements:}
\begin{enumerate*}[(i)]
\item a set of \emph{global states} $\States\subseteq \prod_{i=1}^n L_i$;
\item an \emph{initial state} $\iota \in \States$;
\item a \emph{(partial) global transition function} $T \colon \States\times Act\to\States$
	\st
	$\forall i\in\Agent(a)$, $T(\state_1,a)=\state_2$ iff
	$T_i(\state_1^i,a) = \state^i_2$, whereas $\forall i
	\in\A\setminus\Agent(a)$, $\state_1^i = \state^i_2$, where 
	$\state_1^i$ is the $i$-th local state of $\state_1$; and
\item a \emph{valuation function} $V\colon\States\to2^{\PV}$.
\end{enumerate*}
\end{definition}
\LongVersion{For state $\state = (l_1, \dots, l_n)$, we denote the local component
of agent $i$ by $\state^i=l_i$.
Also, we will sometimes write $\state_1 \trans a \state_2$ instead of $T(\state_1,a) = \state_2$.}

\section{Extending the \AMAS/ model}
\label{sec:EAMAS}

\def\At{\mathit{AT}}
\def\Fun{\mathit{FUN}}
\def\Guards{\mathit{Guards}}
\def\LT{\mathit{LT}}
\def\Exp{\mathit{EXP}}
\def\v{{\bf v}}
\def\Domain{D}
\def\domain{d}

\LongVersion{%
\AMAS/, as defined in~\cite{AAMASWJWPPDAM2018a}, do not include any attribute
feature, which are characteristics of the actions in \ADT/s.  Therefore, we
extend them as \EAMAS/ to take these into account.
This is obtained by extending each local transition with several attributes as
defined below.
}
\ShortVersion{%
As defined in~\cite{AAMASWJWPPDAM2018a}, \AMAS/ do not include attributes.
Therefore, to model \ADT/s we now define \emph{Extended \AMAS/}, associating
attributes to local transitions.
}

\begin{definition}[Extended Asynchronous Multi-Agent Systems]
\label{def:EAMAS}
An \emph{Extended Asynchronous Multi-Agent System (\EAMAS/)} is an \AMAS/ where
each local transition function $t\in\LT=\bigcup_{i\in A} T_{i}$ 
is equipped with a finite set of variables $\At_t= \{v^1_t,\ldots,v^k_t\}$,
called \emph{attributes}, over a domain
$\Domain_t=\domain_t^1\times\cdots\times\domain_t^k$.

Let $\At=\bigcup_{t \in T} \At_t$ and $\Domain=\prod_{t\in T} \Domain_t$.
Let $\Guards$ be the set of formulas of the form
$\beta \sim 0$, where $\beta$ is a linear expression over attributes of $\At$
and ${\sim}\!\in\!\{<,\leq,=,\geq,>\}$.
Let $\Fun$ be all functions taking arguments in $\At$,
and $\Exp(\At,\Fun)$ be linear expressions over $\At$ and $\Fun$.
Each transition \mbox{$t\in\LT$} has associated
\begin{enumerate*}[(i)]
\item \label{def:EAMAS:message}%
	  a \emph{message} $f_M(t)\in (\{!,?\} \times \{\mathit{ok,nok}\}) \cup \{\bot\}$;
\item \label{def:EAMAS:guard}%
	  a \emph{guard} $f_G(t) \in \Guards$; and
\item \label{def:EAMAS:modfun}%
	  a \emph{modification function} $f_t\colon\At_t\to\Exp(\At,\Fun)$.
\end{enumerate*}
\end{definition}

\Cref{def:EAMAS:message} indicates whether $t$ is not a synchronised transition
($\bot$), or whether it is sending (marked with~$!$) or receiving ($?$) a
message $\in\{\mathit{ok,nok}\}$. Guards in \cref{def:EAMAS:guard} constrain
transitions. \Cref{def:EAMAS:modfun} states how to modify the associated
attributes when taking a transition.
To model \ADT/s we must also extend \IIS/.

\begin{definition}[Extended Interleaved Interpreted System]
\label{def:EIIS}
Let $\PV$ be a set of propositional variables,
$\v\colon\At\to\Domain$ a valuation of the attributes,
and $\v_0$ an initial valuation.
An \emph{extended interleaved interpreted system (\EIIS/)} $M$, or a \emph{model},
is an \EAMAS/ extended with\LongVersion{ the following elements:}
\begin{enumerate*}[(i)]
\item \label{def:EIIS:gstates}%
	  a set of \emph{global states}
	  $\States \subseteq L_1\times\dots\times L_n \times \Domain$;
\item \label{def:EIIS:istate}%
	  an \emph{initial state} $s_0=\langle(\iota_1,\dots,\iota_n),\v_0\rangle\in\States$;
\item \label{def:EIIS:gtrans}%
	  a \emph{global transition relation} $T \subseteq \States \times Act \times \States$
	\st
	$\langle(l_1,\ldots,l_n,\v), a, (l'_1,\ldots,l'_n,\v')\rangle \in T$
	iff\LongVersion{ the following conditions are satisfied}:
	\begin{enumerate*}[\bf(a)]
	\item \label{def:EIIS:gtrans:sync}%
		  $\exists i,j \in Agent(a)$ $\land$
		  $\exists t_i = (l_i,a,l'_i) \in T_i$ $\land$
		  $\exists t_j = (l_j,a,l'_j) \in T_j$
	      $\,\st~f_M(t_i)=(!,m) \wedge f_M(t_j)=(?,m)$;\\
	\item \label{def:EIIS:gtrans:async}%
		  $\forall k \in A \setminus \{i,j\}~\,l_k = l'_k$;
	\item \label{def:EIIS:gtrans:enabled}%
		  $\v \models f_G(t_i) \land f_G(t_j)$;
	\item \label{def:EIIS:gtrans:update}%
		  $\v' = \v[\At_{t_i}][\At_{t_j}]$;\\
	and
	\end{enumerate*}
\item \label{def:EIIS:valfun}%
	  a \emph{valuation function} $V: \States \rightarrow 2^{\PV}$,~
\end{enumerate*}
where $\v[\At_{t_i}][\At_{t_j}]$ in \cref{def:EIIS:gtrans:update} indicates the
substitution of attributes in the valuation $\v$ according to transitions $t_i$
and $t_j$, that is $\v' = \v\left[\bigwedge_{v_{t_i} \in \At_{t_i}} v_{t_i} :=
f_{t_i}(v_{t_i})\right]\left[\bigwedge_{v_{t_j} \in \At_{t_j}} v_{t_j} :=
f_{t_j}(v_{t_j})\right]$.
\end{definition}
In the definition of the global transition relation $T$,
\cref{def:EIIS:gtrans:sync} specifies the synchronisation of transition $t_i$
(with a sending action) and $t_j$ (with a receiving action) that share the
message $m$.
\Cref{def:EIIS:gtrans:async} ensures that agents other than $i$ and $j$ do not
change their states in such a synchronisation.
\Cref{def:EIIS:gtrans:enabled} guarantees that the guards of $t_i$ and $t_j$
hold for the valuation $\v$.
Finally, \cref{def:EIIS:gtrans:update} indicates how $\v'$ is obtained by
updating $\v$ with the attributes values modified by $t_i$ and $t_j$.

\LongVersion{For a state $\state = ((l_1, \dots, l_n),\v)$, we denote the local component
of agent $i$ by $\state^i=l_i$.
Also, we will sometimes write $\state_1 \trans a \state_2$ instead of $T(\state_1,a) = \state_2$.}

\section{\EAMAS/ transformation of \ADT/s}
\label{sec:patterns}

\LongVersion{%
In this section we first introduce the transformation patterns from \ADT/s
to \EAMAS/s. Next, we prove that the transformation is correct, \ie{} that
all relevant paths are captured. For that, we take advantage that our
transformation avoids unnecessary interleaving for checking whether a node
action is successful or not, together with the updated values of attributes.
Finally, we apply the transformation patterns on the running example.
}
\ShortVersion{%
Each \ADT/ node is modelled as an \EAMAS/ agent via \emph{transformation
patterns}. These synchronise among themselves via shared actions. The approach
is \emph{compositional}, \ie the resulting \EIIS/ model of the \ADT/ contains an
agent per node in the tree, and \emph{correct}, \ie all relevant (\ADT/) paths
are captured by the model. Note that unlike timed formalisms, \eg (priced) time
automata where clocks let time pass, time in \EAMAS/ is an attribute. This
allows \eg to give minimum and maximum time values, and also makes \EAMAS/ much
easier to model check.
}

\ifdefined\VersionLong
  \subsection{Transformation patterns}
  \label{sec:transformation:patterns}
\else
  \paragraph{Transformation patterns.}
\fi
\LongVersion{%
\Cref{tab:transform:leaf,tab:transform:conjunct,tab:transform:counter,tab:transform:seq}
show how each \ADT/ construct is transformed into an agent (sub)model. Models
communicate via the blue transitions.  Note that attacks and defences in the
agents shown in
\Cref{tab:transform:leaf,tab:transform:conjunct,tab:transform:counter,tab:transform:seq}
operate similarly, both in case of successful and failed operation. The
information on the result is transmitted in both cases, to be taken into account
by the relevant operations at a higher level.  The leaf nodes are presented
in~\Cref{tab:transform:leaf}.
\par
Initially, a leaf either signals action $\actAttack$ (\resp/ defence operation
$\actDefence{}$) or fails. Self-loops in states $l_1, l'_1$ synchronise with all
nodes that depend on this attack (\resp/ defence), which permits handling DAG
\ADT/s.
}
\ShortVersion{%
\Cref{tab:transform:short} shows how each \ADT/ construct is transformed into an
agent (sub)model. Models communicate via the blue transitions. Transformations
are symmetrical for attack and defence nodes: \Cref{tab:transform:short} shows
attack patterns.
A leaf signals action $\actAttack$ or fails. For DAG \ADT/s where gates can
share children, self-loops in states $\loc_1,\loc'_1$ synchronise with \emph{all}
nodes that depend on the attack.
}%
\begin{table}[!h]
\vspace{-2ex}
\ifdefined\VersionLong
	\input{tables/tabTransformLeaf}
	\caption{\ADT/ leaf nodes and corresponding agent models\label{tab:transform:leaf}}
\else

\begin{center}
\scalebox{0.76}{
\begin{tabular}{|c|c|}
\hline
{\bf \ADT/ construct} & {\bf Agent model}\\
\hline\hline
\multicolumn{2}{|c|}{Leaf node}\\
\hline
\parbox[c]{2cm}{\centering
	\vspace{1mm}
	\begin{tikzpicture}[every state/.style={ultra thick,draw=red,
						minimum size=6mm}]
		\node[state] (LA) {$\actAttack{}$};
	\end{tikzpicture}
	\vspace{1mm}
} &
\parbox[c]{8cm}{\centering
	\vspace{1mm}
	\begin{tikzpicture}
			[every state/.style={thick,draw,minimum size=6mm},
				>=stealth',node distance=1.8cm,->]
		\node[state,initial] (l0) {$\loc_0$};
		\node[state,right of=l0] (l1) {$\loc_1$};
		\node[state,below of=l1, node distance=1cm] (l'1) {$\loc'_1$};
		\draw (l0) -- node[above] {$\actAttack{}$} (l1);
		\draw[sync] (l1) to [loop right] node[right]
			{$\styleSync{\send\actAttack{}\_\ok}$} (l1);
		\draw[sync] (l0) -- node[below,sloped]
			{$\styleSync{\send\actAttack{}\_\nok}$} (l'1);
		\draw[sync] (l'1) to [loop right] node[right]
			{$\styleSync{\send\actAttack{}\_\nok}$} (l'1);
	\end{tikzpicture}
	\vspace{1mm}
	}
	\\
\hline\hline
\multicolumn{2}{|c|}{Conjunction/disjunction nodes}\\
\hline
\parbox[c]{4cm}{\centering
  \vspace{1mm}
  \input{./figures/and_attack_gate.tex}\!\!
	\vspace{1mm}
} &
\parbox[c]{10cm}{\centering
	\vspace{1mm}
	\begin{tikzpicture}
			[every state/.style={thick,draw,minimum size=6mm},
				>=stealth',node distance=1.8cm,->]
		\node[state,initial] (l0) {$\loc_0$};
		\node[state,right of=l0] (l1) {$\loc_1$};
		\node[state,right of=l1] (l2) {$\loc_2$};
		\node[draw=none,right of=l2,node distance=0.9cm] {$\cdots$};
		\node[state,right of=l2] (l3) {$\loc_n$};
		\node[state,right of=l3] (l4) {$\loc_\ActAttack{}$};
		\node[state,below of=l2] (l'1) {$\loc'_1$};
		\draw[sync] (l0) --
			node[above] {$\styleSync{\receive\actAttack{}_1\_\ok}$} (l1);
		\draw[sync] (l1) --
			node[above] {$\styleSync{\receive\actAttack{}_2\_\ok}$} (l2);
		\draw (l3) -- node[above] {$\ActAttack{}$} (l4);
		\draw[sync] (l4) to [loop right] node[right] {$\styleSync{\send\ActAttack{}\_\ok}$}
		(l4);
		\draw[sync] (l0) -- node[above,sloped]
			{$\styleSync{\receive\actAttack{}_1\_\nok}$} (l'1);
		\draw[sync] (l0) to [bend right=20] node[above,sloped]
			{$\styleSync{\receive\actAttack{}_2\_\nok}$} (l'1);
		\draw[sync] (l0) to [bend right=45] node[above,sloped] {$\vdots$}
			node[below,sloped]
			{$\styleSync{\receive\actAttack{}_n\_\nok}$} (l'1);
		\draw[sync] (l'1) to [loop right]
			node[right] {$\styleSync{\send\ActAttack{}\_\nok}$} (l'1);
	\end{tikzpicture}

	\vspace{1mm}
	}
	\\
\hline
\parbox[c]{4cm}{\centering
  \vspace{1mm}
  \input{./figures/or_attack_gate.tex}\!\!
	\vspace{1mm}
} &
\parbox[c]{10cm}{\centering
	\vspace{1mm}
	\begin{tikzpicture}
			[every state/.style={thick,draw,minimum size=6mm},
				>=stealth',node distance=1.8cm,->]
		\node[state,initial] (l0) {$\loc_0$};
		\node[state,right of=l0] (l1) {$\loc_1$};
		\node[state,right of=l1] (l2) {$\loc_\ActAttack{}$};
		\node[draw=none,right of=l0,node distance=0.9cm,yshift=0.1cm] {$\vdots$};
		\node[state,below of=l1] (l'1) {$\loc'_1$};
		\node[state,right of=l'1] (l'2) {$\loc'_2$};
		\node[draw=none,right of=l'2,node distance=0.9cm] {$\cdots$};
		\node[state,right of=l'2] (l'3) {$\loc'_n$};
		\draw[sync] (l0) to [bend left]
			node[above] {$\styleSync{\receive\actAttack{}_1\_\ok}$} (l1);
		\draw[sync] (l0) to [bend right]
			node[below] {$\styleSync{\receive\actAttack{}_n\_\ok}$} (l1);
		\draw (l1) -- node[above] {$\ActAttack{}$} (l2);
		\draw[sync] (l2) to [loop right]
			node[right] {$\styleSync{\send\ActAttack{}\_\ok}$} (l2);
		\draw[sync] (l0) to [bend right=20] node[below,sloped]
			{$\styleSync{\receive\actAttack{}_1\_\nok}$} (l'1);
		\draw[sync] (l'1) -- node[above]
			{$\styleSync{\receive\actAttack{}_2\_\nok}$} (l'2);
		\draw[sync] (l'3) to [loop right]
			node[right] {$\styleSync{\send\ActAttack{}\_\nok}$} (l'3);
	\end{tikzpicture}

	\vspace{1mm}
	}
	\\
\hline\hline
\multicolumn{2}{|c|}{Countering nodes}\\
\hline
\parbox[c]{4cm}{\centering
  \vspace{1mm}
  \input{./figures/counter_defence_gate.tex}\!\!
	\vspace{1mm}
} &
\parbox[c]{10cm}{\centering
	\vspace{1mm}
	\begin{tikzpicture}
			[every state/.style={thick,draw,minimum size=6mm},
				>=stealth',node distance=1.8cm,->]
		\node[state,initial] (l0) {$\loc_0$};
		\node[state,right of=l0] (l1) {$\loc_1$};
		\node[state,below of=l1] (l'1) {$\loc'_1$};
		\node[state,right of=l1] (l2) {$\loc_2$};
		\node[state,right of=l2] (lA) {$\loc_A$};
		\draw[sync] (l0) --
			node[above] {$\styleSync{\receive\actAttack{}\_\ok}$} (l1);
		\draw[sync] (l1) --
			node[above] {$\styleSync{\receive\actDefence{}\_\nok}$} (l2);
		\draw[sync] (l0) -- node[above,sloped]
			{$\styleSync{\receive\actAttack{}\_\nok}$} (l'1);
		\draw[sync] (l0) to [bend right] node[below,sloped]
			{$\styleSync{\receive\actDefence{}\_\ok}$} (l'1);
		\draw (l2) -- node[above] {$\ActAttack{}$} (lA);
		\draw[sync] (lA) to [loop right]
			node[right] {$\styleSync{\send\ActAttack{}\_\ok}$} (lA);
		\draw[sync] (l'1) to [loop right]
			node[right] {$\styleSync{\send\ActAttack{}\_\nok}$} (l'1);
	\end{tikzpicture}

	\vspace{1mm}
	}
	\\
\hline
\parbox[c]{4cm}{\centering
  \vspace{1mm}
  \input{./figures/no_defence_gate.tex}
	\vspace{1mm}
} &
\parbox[c]{10cm}{\centering
	\vspace{1mm}
	\begin{tikzpicture}
			[every state/.style={thick,draw,minimum size=6mm},
				>=stealth',node distance=1.8cm,->]
		\node[state,initial] (l0) {$\loc_0$};
		\node[state,right of=l0] (l1) {$\loc_1$};
		\node[state,right of=l1] (l2) {$\loc_\ActAttack{}$};
		\node[state,below of=l1] (l'1) {$\loc'_1$};
		\node[state,right of=l'1] (l'2) {$\loc'_2$};
		\draw[sync] (l0) to [bend left]
			node[above] {$\styleSync{\receive\actAttack{}\_\ok}$} (l1);
		\draw[sync] (l0) to [bend right]
			node[below] {$\styleSync{\receive\actDefence{}\_\nok}$} (l1);
		\draw (l1) -- node[above] {$\ActAttack{}$} (l2);
		\draw[sync] (l2) to [loop right]
			node[right] {$\styleSync{\send\ActAttack{}\_\ok}$} (l2);
		\draw[sync] (l0) to [bend right=20] node[below,sloped]
			{$\styleSync{\receive\actAttack{}\_\nok}$} (l'1);
		\draw[sync] (l'1) -- node[above]
			{$\styleSync{\receive\actDefence{}\_\ok}$} (l'2);
		\draw[sync] (l'2) to [loop right]
			node[right] {$\styleSync{\send\ActAttack{}\_\nok}$} (l'2);
	\end{tikzpicture}

	\vspace{1mm}
	}
	\\
\hline\hline
\multicolumn{2}{|c|}{Sequential nodes}\\
\hline
\parbox[c]{4cm}{\centering
  \vspace{1mm}
  \input{./figures/sand_attack_gate.tex}
	\vspace{1mm}
} &
\parbox[c]{10cm}{\centering
	\vspace{1mm}
	\begin{tikzpicture}
			[every state/.style={thick,draw,minimum size=6mm},
				>=stealth',node distance=1.8cm,->]
		\node[state,initial] (l0) {$\loc_0$};
		\node[state,right of=l0] (l1) {$\loc_1$};
		\node[state,below of=l1] (l'1) {$\loc'_1$};
		\node[state,right of=l1] (l2) {$\loc_2$};
		\node[draw=none,right of=l2,node distance=0.9cm] {$\cdots$};
		\node[state,right of=l2] (ln) {$\loc_n$};
		\node[state,right of=ln] (lA) {$\loc_A$};
		\draw[sync] (l0) --
			node[above] {$\styleSync{\receive\actAttack{}_1\_\ok}$} (l1);
		\draw[sync] (l1) --
			node[above] {$\styleSync{\receive\actAttack{}_2\_\ok}$} (l2);
		\draw[sync] (l0) -- node[above,sloped]
			{$\styleSync{\receive\actAttack{}_1\_\nok}$} (l'1);
		\draw[sync] (l1) to node[above,sloped]
			{$\styleSync{\receive\actAttack{}_2\_\nok}$} (l'1);
		\draw[sync] (l2) to node[above,sloped]
			{$\styleSync{\receive\actAttack{}_3\_\nok}$} (l'1);
		\draw (ln) -- node[above] {$\ActAttack{}$} (lA);
		\draw[sync] (lA) to [loop right]
			node[right] {$\styleSync{\send\ActAttack{}\_\ok}$} (lA);
		\draw[sync] (l'1) to [loop right]
			node[right] {$\styleSync{\send\ActAttack{}\_\nok}$} (l'1);
	\end{tikzpicture}
	\vspace{1mm}
	}
  \\
\hline
\parbox[c]{4cm}{\centering
  \vspace{1mm}
  \input{./figures/snand_attack_gate.tex}
	\vspace{1mm}
} &
\parbox[c]{10cm}{\centering
	\vspace{1mm}
	\begin{tikzpicture}
			[every state/.style={thick,draw,minimum size=6mm},
				>=stealth',node distance=1.8cm,->]
		\node[state,initial] (l0) {$\loc_0$};
		\node[state,right of=l0] (l1) {$\loc_1$};
		\node[state,below of=l1] (l'1) {$\loc'_1$};
		\node[state,right of=l1] (l2) {$\loc_2$};
		\node[state,right of=l2] (lA) {$\loc_A$};
		\draw[sync] (l0) --
			node[above] {$\styleSync{\receive\actAttack{}\_\ok}$} (l1);
		\draw[sync] (l1) --
			node[above] {$\styleSync{\receive\actDefence{}\_\nok}$} (l2);
		\draw[sync] (l0) -- node[above,sloped]
			{$\styleSync{\receive\actAttack{}\_\nok}$} (l'1);
		\draw[sync] (l1) to node[above,sloped]
			{$\styleSync{\receive\actDefence{}\_\ok}$} (l'1);
		\draw (l2) -- node[above] {$\ActAttack{}$} (lA);
		\draw[sync] (lA) to [loop right]
			node[right] {$\styleSync{\send\ActAttack{}\_\ok}$} (lA);
		\draw[sync] (l'1) to [loop right]
			node[right] {$\styleSync{\send\ActAttack{}\_\nok}$} (l'1);
	\end{tikzpicture}
	\vspace{1mm}
	}
  \\
\hline
\end{tabular}
}
\end{center}
	\caption{\ADT/ nodes and corresponding agent models\label{tab:transform:short}}
\fi
\vspace{-2ex}
\end{table}
\LongVersion{%
The conjunction and disjunction operators, concerning either attacks or
defences, are described in~\Cref{tab:transform:conjunct}.  For conjunctions,
\emph{any} action can be signaled, and then another, until all are. Then, the
actual attack $\ActAttack{}$ (\resp/ defence $\ActDefence{}$) occurs, after
which the model signals a completed operation. If an action fails, the
conjunction fails immediately after. Disjunctions and counter actions operate in
the expected analogous way.
}%
\ShortVersion{%
In conjunctions, (all) actions must be signaled in \emph{any} order. Then, attack
$\ActAttack$ occurs, followed by the model signaling success. The conjunction
fails if any action fails. Disjunctions and counter actions operate in the
expected analogous way.
Sequential constructs enforce the absence of parallelism, as per the semantics
chosen for \gateSAND and \gateSNAND gates (see~\Cref{sec:adt:basic}).
}%
Note that patterns use a single order of actions: other orders
are shufflings of the semantically equivalent sequence shown in
\LongVersion{\Cref{tab:transform:leaf}}%
\ShortVersion{\Cref{tab:transform:short}}%
, and are thus already present in the state space.
\lp{related to PORs?}
\LongVersion{%
\begin{table}[!!htb]

\begin{center}
\scalebox{0.9}{
\begin{tabular}{|c|c|}
\hline\hline
{\bf \ADT/ construct} & {\bf Agent model}\\
\hline\hline
\parbox[c]{4cm}{\centering
	\vspace{1mm}
	\begin{tikzpicture}
			[every node/.style={ultra thick,draw=red,minimum size=6mm}]
		\node[and gate US,point up,logic gate inputs=nn] (A)
			{\rotatebox{-90}{$\ActAttack$}};
		\node[state] at (-1,-1.3) (a1) {$\actAttack_1$};
		\node[state] at (1,-1.3) (an) {$\actAttack_n$};
		\node[draw=none] at (0,-1.3) {$\cdots$};
		\draw (a1.north) -- ([yshift=0.15cm]a1.north) -| (A.input 1);
		\draw (an.north) -- ([yshift=0.15cm]an.north) -| (A.input 2);
	\end{tikzpicture}

	\vspace{1mm}
	}
	& \parbox[c]{10cm}{\centering
	\vspace{1mm}
	\begin{tikzpicture}
			[every state/.style={thick,draw,minimum size=6mm},
				>=stealth',node distance=1.8cm,->]
		\node[state,initial] (l0) {$\loc_0$};
		\node[state,right of=l0] (l1) {$\loc_1$};
		\node[state,right of=l1] (l2) {$\loc_2$};
		\node[draw=none,right of=l2,node distance=0.9cm] {$\cdots$};
		\node[state,right of=l2] (l3) {$\loc_n$};
		\node[state,right of=l3] (l4) {$\loc_\ActAttack{}$};
		\node[state,below of=l2] (l'1) {$\loc'_1$};
		\draw[sync] (l0) --
			node[above] {$\styleSync{\receive\actAttack{}_1\_\ok}$} (l1);
		\draw[sync] (l1) --
			node[above] {$\styleSync{\receive\actAttack{}_2\_\ok}$} (l2);
		\draw (l3) -- node[above] {$\ActAttack{}$} (l4);
		\draw[sync] (l4) to [loop right] node[right] {$\styleSync{\send\ActAttack{}\_\ok}$}
		(l4);
		\draw[sync] (l0) -- node[above,sloped]
			{$\styleSync{\receive\actAttack{}_1\_\nok}$} (l'1);
		\draw[sync] (l0) to [bend right=20] node[above,sloped]
			{$\styleSync{\receive\actAttack{}_2\_\nok}$} (l'1);
		\draw[sync] (l0) to [bend right=45] node[above,sloped] {$\vdots$}
			node[below,sloped]
			{$\styleSync{\receive\actAttack{}_n\_\nok}$} (l'1);
		\draw[sync] (l'1) to [loop right]
			node[right] {$\styleSync{\send\ActAttack{}\_\nok}$} (l'1);
	\end{tikzpicture}

	\vspace{1mm}
	}
	\\
\hline
\parbox[c]{4cm}{\centering
	\vspace{1mm}
	\begin{tikzpicture}
			[every node/.style={ultra thick,draw=Green,minimum size=6mm},
				every state/.style={rectangle}]
		\node[and gate US,point up,logic gate inputs=nn] (D)
			{\rotatebox{-90}{$\ActDefence$}};
		\node[state] at (-1,-1.3) (d1) {$\actDefence_1$};
		\node[state] at (1,-1.3) (dn) {$\actDefence_n$};
		\node[draw=none] at (0,-1.3) {$\cdots$};
		\draw (d1.north) -- ([yshift=0.15cm]d1.north) -| (D.input 1);
		\draw (dn.north) -- ([yshift=0.15cm]dn.north) -| (D.input 2);
	\end{tikzpicture}

	\vspace{1mm}
	}
	& \parbox[c]{10cm}{\centering
	\vspace{1mm}
	\begin{tikzpicture}
			[every state/.style={thick,draw,minimum size=6mm},
				>=stealth',node distance=1.8cm,->]
		\node[state,initial] (l0) {$\loc_0$};
		\node[state,right of=l0] (l1) {$\loc_1$};
		\node[state,right of=l1] (l2) {$\loc_2$};
		\node[draw=none,right of=l2,node distance=0.9cm] {$\cdots$};
		\node[state,right of=l2] (l3) {$\loc_n$};
		\node[state,right of=l3] (l4) {$\loc_\ActDefence{}$};
		\node[state,below of=l2] (l'1) {$\loc'_1$};
		\draw[sync] (l0) -- node[above]
			{$\styleSync{\receive\actDefence{}_1\_\ok}$} (l1);
		\draw[sync] (l1) -- node[above]
			{$\styleSync{\receive\actDefence{}_2\_\ok}$} (l2);
		\draw (l3) -- node[above] {$\ActDefence{}$} (l4);
		\draw[sync] (l4) to [loop right]
			node[right] {$\styleSync{\send\ActDefence{}\_\ok}$} (l4);
		\draw[sync] (l0) -- node[above,sloped]
			{$\styleSync{\receive\actDefence{}_1\_\nok}$} (l'1);
		\draw[sync] (l0) to [bend right=20] node[above,sloped]
			{$\styleSync{\receive\actDefence{}_2\_\nok}$} (l'1);
		\draw[sync] (l0) to [bend right=45] node[above,sloped] {$\vdots$}
			node[below,sloped]
			{$\styleSync{\receive\actDefence{}_n\_\nok}$} (l'1);
		\draw[sync] (l'1) to [loop right]
			node[right] {$\styleSync{\send\ActDefence{}\_\nok}$} (l'1);
	\end{tikzpicture}

	\vspace{1mm}
	}
	\\
\hline
\parbox[c]{4cm}{\centering
	\vspace{1mm}
	\begin{tikzpicture}
			[every node/.style={ultra thick,draw=red,minimum size=6mm}]
		\node[or gate US,point up,logic gate inputs=nn] (A)
			{\rotatebox{-90}{$\ActAttack$}};
		\node[state] at (-1,-1.3) (a1) {$\actAttack_1$};
		\node[state] at (1,-1.3) (an) {$\actAttack_n$};
		\node[draw=none] at (0,-1.3) {$\cdots$};
		\draw (a1.north) -- ([yshift=0.15cm]a1.north) -| (A.input 1);
		\draw (an.north) -- ([yshift=0.15cm]an.north) -| (A.input 2);
	\end{tikzpicture}

	\vspace{1mm}
	}
	& \parbox[c]{10cm}{\centering
	\vspace{1mm}
	\begin{tikzpicture}
			[every state/.style={thick,draw,minimum size=6mm},
				>=stealth',node distance=1.8cm,->]
		\node[state,initial] (l0) {$\loc_0$};
		\node[state,right of=l0] (l1) {$\loc_1$};
		\node[state,right of=l1] (l2) {$\loc_\ActAttack{}$};
		\node[draw=none,right of=l0,node distance=0.9cm,yshift=0.1cm] {$\vdots$};
		\node[state,below of=l1] (l'1) {$\loc'_1$};
		\node[state,right of=l'1] (l'2) {$\loc'_2$};
		\node[draw=none,right of=l'2,node distance=0.9cm] {$\cdots$};
		\node[state,right of=l'2] (l'3) {$\loc'_n$};
		\draw[sync] (l0) to [bend left]
			node[above] {$\styleSync{\receive\actAttack{}_1\_\ok}$} (l1);
		\draw[sync] (l0) to [bend right]
			node[below] {$\styleSync{\receive\actAttack{}_n\_\ok}$} (l1);
		\draw (l1) -- node[above] {$\ActAttack{}$} (l2);
		\draw[sync] (l2) to [loop right]
			node[right] {$\styleSync{\send\ActAttack{}\_\ok}$} (l2);
		\draw[sync] (l0) to [bend right=20] node[below,sloped]
			{$\styleSync{\receive\actAttack{}_1\_\nok}$} (l'1);
		\draw[sync] (l'1) -- node[above]
			{$\styleSync{\receive\actAttack{}_2\_\nok}$} (l'2);
		\draw[sync] (l'3) to [loop right]
			node[right] {$\styleSync{\send\ActAttack{}\_\nok}$} (l'3);
	\end{tikzpicture}

	\vspace{1mm}
	}
	\\
\hline
\parbox[c]{4cm}{\centering
	\vspace{1mm}
	\begin{tikzpicture}
			[every node/.style={ultra thick,draw=Green,minimum size=6mm},
				every state/.style={rectangle}]
		\node[or gate US,point up,logic gate inputs=nn] (D)
			{\rotatebox{-90}{$\ActDefence$}};
		\node[state] at (-1,-1.3) (d1) {$\actDefence_1$};
		\node[state] at (1,-1.3) (dn) {$\actDefence_n$};
		\node[draw=none] at (0,-1.3) {$\cdots$};
		\draw (d1.north) -- ([yshift=0.15cm]d1.north) -| (D.input 1);
		\draw (dn.north) -- ([yshift=0.15cm]dn.north) -| (D.input 2);
	\end{tikzpicture}

	\vspace{1mm}
	}
	& \parbox[c]{10cm}{\centering
	\vspace{1mm}
	\begin{tikzpicture}
			[every state/.style={thick,draw,minimum size=6mm},
				>=stealth',node distance=1.8cm,->]
		\node[state,initial] (l0) {$\loc_0$};
		\node[state,right of=l0] (l1) {$\loc_1$};
		\node[state,right of=l1] (l2) {$\loc_\ActDefence{}$};
		\node[draw=none,right of=l0,node distance=0.9cm,yshift=0.1cm] {$\vdots$};
		\node[state,below of=l1] (l'1) {$\loc'_1$};
		\node[state,right of=l'1] (l'2) {$\loc'_2$};
		\node[draw=none,right of=l'2,node distance=0.9cm] {$\cdots$};
		\node[state,right of=l'2] (l'3) {$\loc'_n$};
		\draw[sync] (l0) to [bend left]
			node[above] {$\styleSync{\receive\actDefence{}_1\_\ok}$} (l1);
		\draw[sync] (l0) to [bend right]
			node[below] {$\styleSync{\receive\actDefence{}_n\_\ok}$} (l1);
		\draw (l1) -- node[above] {$\ActDefence{}$} (l2);
		\draw[sync] (l2) to [loop right]
			node[right] {$\styleSync{\send\ActDefence{}\_\ok}$} (l2);
		\draw[sync] (l0) to [bend right=20] node[below,sloped]
			{$\styleSync{\receive\actDefence{}_1\_\nok}$} (l'1);
		\draw[sync] (l'1) -- node[above]
			{$\styleSync{\receive\actDefence{}_2\_\nok}$} (l'2);
		\draw[sync] (l'3) to [loop right]
			node[right] {$\styleSync{\send\ActDefence{}\_\nok}$} (l'3);
	\end{tikzpicture}

	\vspace{1mm}
	}
	\\
\hline\hline
\end{tabular}
}
\end{center}
	\caption{\ADT/ conjunction/disjunction nodes and corresponding agent models}
	\label{tab:transform:conjunct}
\end{table}
\Cref{tab:transform:counter} shows the models for countering defences or
attacks.
\begin{table}[!!htb]

\begin{center}
\scalebox{0.9}{
\begin{tabular}{|c|c|}
\hline\hline
{\bf \ADT/ construct} & {\bf Agent model}\\
\hline\hline
\parbox[c]{4cm}{\centering
	\vspace{1mm}
	\begin{tikzpicture}
			[every node/.style={ultra thick,draw=red,minimum size=6mm}]
		\node[and gate US,point up,logic gate inputs=ni] (A)
			{\rotatebox{-90}{\large$\ActAttack$}};
		\node[state] at (-1,-1.3) (a) {$\actAttack$};
		\node[rectangle,draw=Green] at (1,-1.3) (d) {$\actDefence$};
		\draw (a.north) -- ([yshift=0.15cm]a.north) -| (A.input 1);
		\draw (d.north) -- ([yshift=0.24cm]d.north) -| (A.input 2);
	\end{tikzpicture}

	\vspace{1mm}
	}
	& \parbox[c]{10cm}{\centering
	\vspace{1mm}
	\begin{tikzpicture}
			[every state/.style={thick,draw,minimum size=6mm},
				>=stealth',node distance=1.8cm,->]
		\node[state,initial] (l0) {$\loc_0$};
		\node[state,right of=l0] (l1) {$\loc_1$};
		\node[state,below of=l1] (l'1) {$\loc'_1$};
		\node[state,right of=l1] (l2) {$\loc_2$};
		\node[state,right of=l2] (lA) {$\loc_A$};
		\draw[sync] (l0) --
			node[above] {$\styleSync{\receive\actAttack{}\_\ok}$} (l1);
		\draw[sync] (l1) --
			node[above] {$\styleSync{\receive\actDefence{}\_\nok}$} (l2);
		\draw[sync] (l0) -- node[above,sloped]
			{$\styleSync{\receive\actAttack{}\_\nok}$} (l'1);
		\draw[sync] (l0) to [bend right] node[below,sloped]
			{$\styleSync{\receive\actDefence{}\_\ok}$} (l'1);
		\draw (l2) -- node[above] {$\ActAttack{}$} (lA);
		\draw[sync] (lA) to [loop right]
			node[right] {$\styleSync{\send\ActAttack{}\_\ok}$} (lA);
		\draw[sync] (l'1) to [loop right]
			node[right] {$\styleSync{\send\ActAttack{}\_\nok}$} (l'1);
	\end{tikzpicture}

	\vspace{1mm}
	}
	\\
\hline
\parbox[c]{4cm}{\centering
	\vspace{1mm}
	\begin{tikzpicture}
			[every node/.style={ultra thick,draw=Green,minimum size=6mm}]
		\node[and gate US,point up,logic gate inputs=ni] (D)
			{\rotatebox{-90}{\large$\ActDefence$}};
		\node[rectangle] at (-1,-1.3) (d) {$\actDefence$};
		\node[state,draw=red] at (1,-1.3) (a) {$\actAttack$};
		\draw (d.north) -- ([yshift=0.15cm]d.north) -| (D.input 1);
		\draw (a.north) -- ([yshift=0.2cm]a.north) -| (D.input 2);
	\end{tikzpicture}

	\vspace{1mm}
	}
	& \parbox[c]{10cm}{\centering
	\vspace{1mm}
	\begin{tikzpicture}
			[every state/.style={thick,draw,minimum size=6mm},
				>=stealth',node distance=1.8cm,->]
		\node[state,initial] (l0) {$\loc_0$};
		\node[state,right of=l0] (l1) {$\loc_1$};
		\node[state,below of=l1] (l'1) {$\loc'_1$};
		\node[state,right of=l1] (l2) {$\loc_2$};
		\node[state,right of=l2] (lD) {$\loc_D$};
		\draw[sync] (l0) --
			node[above] {$\styleSync{\receive\actDefence{}\_\ok}$} (l1);
		\draw[sync] (l1) --
			node[above] {$\styleSync{\receive\actAttack{}\_\nok}$} (l2);
		\draw[sync] (l0) -- node[above,sloped]
			{$\styleSync{\receive\actDefence{}\_\nok}$} (l'1);
		\draw[sync] (l0) to [bend right] node[below,sloped]
			{$\styleSync{\receive\actAttack{}\_\ok}$} (l'1);
		\draw (l2) -- node[above] {$\ActDefence{}$} (lD);
		\draw[sync] (lD) to [loop right]
			node[right] {$\styleSync{\send\ActDefence{}\_\ok}$} (lD);
		\draw[sync] (l'1) to [loop right]
			node[right] {$\styleSync{\send\ActDefence{}\_\nok}$} (l'1);
	\end{tikzpicture}

	\vspace{1mm}
	}
	\\
\hline
\parbox[c]{4cm}{\centering
	\vspace{1mm}
	\begin{tikzpicture}
			[every node/.style={ultra thick,draw=red,minimum size=6mm}]
		\node[or gate US,point up,logic gate inputs=ni] (A)
			{\rotatebox{-90}{\large$\ActAttack$}};
		\node[state] at (-1,-1.3) (a) {$\actAttack$};
		\node[rectangle,draw=Green] at (1,-1.3) (d) {$\actDefence$};
		\draw (a.north) -- ([yshift=0.15cm]a.north) -| (A.input 1);
		\draw (d.north) -- ([yshift=0.24cm]d.north) -| (A.input 2);
	\end{tikzpicture}

	\vspace{1mm}
	}
	& \parbox[c]{10cm}{\centering
	\vspace{1mm}
	\begin{tikzpicture}
			[every state/.style={thick,draw,minimum size=6mm},
				>=stealth',node distance=1.8cm,->]
		\node[state,initial] (l0) {$\loc_0$};
		\node[state,right of=l0] (l1) {$\loc_1$};
		\node[state,right of=l1] (l2) {$\loc_\ActAttack{}$};
		\node[state,below of=l1] (l'1) {$\loc'_1$};
		\node[state,right of=l'1] (l'2) {$\loc'_2$};
		\draw[sync] (l0) to [bend left]
			node[above] {$\styleSync{\receive\actAttack{}\_\ok}$} (l1);
		\draw[sync] (l0) to [bend right]
			node[below] {$\styleSync{\receive\actDefence{}\_\nok}$} (l1);
		\draw (l1) -- node[above] {$\ActAttack{}$} (l2);
		\draw[sync] (l2) to [loop right]
			node[right] {$\styleSync{\send\ActAttack{}\_\ok}$} (l2);
		\draw[sync] (l0) to [bend right=20] node[below,sloped]
			{$\styleSync{\receive\actAttack{}\_\nok}$} (l'1);
		\draw[sync] (l'1) -- node[above]
			{$\styleSync{\receive\actDefence{}\_\ok}$} (l'2);
		\draw[sync] (l'2) to [loop right]
			node[right] {$\styleSync{\send\ActAttack{}\_\nok}$} (l'2);
	\end{tikzpicture}

	\vspace{1mm}
	}
	\\
\hline
\parbox[c]{4cm}{\centering
	\vspace{1mm}
	\begin{tikzpicture}
			[every node/.style={ultra thick,draw=Green,minimum size=6mm}]
		\node[or gate US,point up,logic gate inputs=ni] (D)
			{\rotatebox{-90}{\large$\ActDefence$}};
		\node[rectangle] at (-1,-1.3) (d) {$\actDefence$};
		\node[state,draw=red] at (1,-1.3) (a) {$\actAttack$};
		\draw (d.north) -- ([yshift=0.15cm]d.north) -| (D.input 1);
		\draw (a.north) -- ([yshift=0.2cm]a.north) -| (D.input 2);
	\end{tikzpicture}

	\vspace{1mm}
	}
	& \parbox[c]{10cm}{\centering
	\vspace{1mm}
	\begin{tikzpicture}
			[every state/.style={thick,draw,minimum size=6mm},
				>=stealth',node distance=1.8cm,->]
		\node[state,initial] (l0) {$\loc_0$};
		\node[state,right of=l0] (l1) {$\loc_1$};
		\node[state,right of=l1] (l2) {$\loc_\ActDefence{}$};
		\node[state,below of=l1] (l'1) {$\loc'_1$};
		\node[state,right of=l'1] (l'2) {$\loc'_2$};
		\draw[sync] (l0) to [bend left]
			node[above] {$\styleSync{\receive\actDefence{}\_\ok}$} (l1);
		\draw[sync] (l0) to [bend right]
			node[below] {$\styleSync{\receive\actAttack{}\_\nok}$} (l1);
		\draw (l1) -- node[above] {$\ActDefence{}$} (l2);
		\draw[sync] (l2) to [loop right]
			node[right] {$\styleSync{\send\ActDefence{}\_\ok}$} (l2);
		\draw[sync] (l0) to [bend right=20] node[below,sloped]
			{$\styleSync{\receive\actDefence{}\_\nok}$} (l'1);
		\draw[sync] (l'1) -- node[above]
			{$\styleSync{\receive\actAttack{}\_\ok}$} (l'2);
		\draw[sync] (l'2) to [loop right]
			node[right] {$\styleSync{\send\ActDefence{}\_\nok}$} (l'2);
	\end{tikzpicture}

	\vspace{1mm}
	}
	\\
\hline\hline
\end{tabular}
}
\end{center}
	\caption{\ADT/ countering nodes and corresponding agent models\label{tab:transform:counter}}
\end{table}
Finally, \Cref{tab:transform:seq} shows the models for sequential gates  It is
similar to the Priced Timed Automata in~\cite[Sec.~4.1]{KumarRS15}, which can be
seen as the particular case $n=2$ of the automata in \Cref{tab:transform:seq}.
\begin{table}[!!htb]

\begin{center}
\scalebox{0.9}{
\begin{tabular}{|c|c|}
\hline\hline
{\bf \ADT/ construct} & {\bf Agent model}\\
\hline\hline
\parbox[c]{4cm}{\centering
  \vspace{1mm}

\begingroup
\def\nodesubtree{\node[isosceles triangle, isosceles triangle apex angle=70, shape border rotate=90, minimum size=6.5mm]}

\begin{tikzpicture}
    [every node/.style={ultra thick,draw=red,minimum size=6mm}]
	\normalsize
	\node[and gate US,point up,logic gate inputs=nn, seq=4pt] (A)
		{\rotatebox{-90}{\large$\ActAttack$}};
	\nodesubtree at (-1,-1.3) (a1) {$\!\actAttack_1\!\!$};
	\nodesubtree at ( 1,-1.3) (an) {$\!\actAttack_n\!\!$};
	\node[draw=none] at (0,-1.3) {$\cdots$};
	\draw (a1.north) -- ([yshift=1.5mm]a1.north) -| (A.input 1);
	\draw (an.north) -- ([yshift=1.5mm]an.north) -| (A.input 2);
\end{tikzpicture}
\endgroup
	\vspace{1mm}
	}
& \parbox[c]{10cm}{\centering
	\vspace{1mm}
	\begin{tikzpicture}
			[every state/.style={thick,draw,minimum size=6mm},
				>=stealth',node distance=1.8cm,->]
		\node[state,initial] (l0) {$\loc_0$};
		\node[state,right of=l0] (l1) {$\loc_1$};
		\node[state,below of=l1] (l'1) {$\loc'_1$};
		\node[state,right of=l1] (l2) {$\loc_2$};
		\node[draw=none,right of=l2,node distance=0.9cm] {$\cdots$};
		\node[state,right of=l2] (ln) {$\loc_n$};
		\node[state,right of=ln] (lA) {$\loc_A$};
		\draw[sync] (l0) --
			node[above] {$\styleSync{\receive\actAttack{}_1\_\ok}$} (l1);
		\draw[sync] (l1) --
			node[above] {$\styleSync{\receive\actAttack{}_2\_\ok}$} (l2);
		\draw[sync] (l0) -- node[above,sloped]
			{$\styleSync{\receive\actAttack{}_1\_\nok}$} (l'1);
		\draw[sync] (l1) to node[above,sloped]
			{$\styleSync{\receive\actAttack{}_2\_\nok}$} (l'1);
		\draw[sync] (l2) to node[above,sloped]
			{$\styleSync{\receive\actAttack{}_3\_\nok}$} (l'1);
		\draw (ln) -- node[above] {$\ActAttack{}$} (lA);
		\draw[sync] (lA) to [loop right]
			node[right] {$\styleSync{\send\ActAttack{}\_\ok}$} (lA);
		\draw[sync] (l'1) to [loop right]
			node[right] {$\styleSync{\send\ActAttack{}\_\nok}$} (l'1);
	\end{tikzpicture}
	\vspace{1mm}
	}
  \\
\hline

\parbox[c]{4cm}{\centering
  \vspace{1mm}

\begingroup
\def\nodesubtree{\node[isosceles triangle, isosceles triangle apex angle=70, shape border rotate=90, minimum size=6.5mm]}

\begin{tikzpicture}
	[every node/.style={ultra thick,draw=Green,minimum size=6mm},every state/.style={rectangle}]
	\normalsize
	\node[and gate US,point up,logic gate inputs=nn, seq=4pt] (D)
		{\rotatebox{-90}{\large$\ActDefence$}};
	\nodesubtree at (-1,-1.3) (d1) {$\!\actDefence_1\!\!$};
	\nodesubtree at ( 1,-1.3) (dn) {$\!\actDefence_n\!\!$};
	\node[draw=none] at (0,-1.3) {$\cdots$};
	\draw (d1.north) -- ([yshift=1.5mm]d1.north) -| (D.input 1);
	\draw (dn.north) -- ([yshift=1.5mm]dn.north) -| (D.input 2);
\end{tikzpicture}
\endgroup
	\vspace{1mm}
	}
& \parbox[c]{10cm}{\centering
	\vspace{1mm}
	\begin{tikzpicture}
			[every state/.style={thick,draw,minimum size=6mm},
				>=stealth',node distance=1.8cm,->]
		\node[state,initial] (l0) {$\loc_0$};
		\node[state,right of=l0] (l1) {$\loc_1$};
		\node[state,below of=l1] (l'1) {$\loc'_1$};
		\node[state,right of=l1] (l2) {$\loc_2$};
		\node[draw=none,right of=l2,node distance=0.9cm] {$\cdots$};
		\node[state,right of=l2] (ln) {$\loc_n$};
		\node[state,right of=ln] (lD) {$\loc_D$};
		\draw[sync] (l0) --
			node[above] {$\styleSync{\receive\actDefence{}_1\_\ok}$} (l1);
		\draw[sync] (l1) --
			node[above] {$\styleSync{\receive\actDefence{}_2\_\ok}$} (l2);
		\draw[sync] (l0) -- node[above,sloped]
			{$\styleSync{\receive\actDefence{}_1\_\nok}$} (l'1);
		\draw[sync] (l1) to node[above,sloped]
			{$\styleSync{\receive\actDefence{}_2\_\nok}$} (l'1);
		\draw[sync] (l2) to node[above,sloped]
			{$\styleSync{\receive\actDefence{}_3\_\nok}$} (l'1);
		\draw (ln) -- node[above] {$\ActDefence{}$} (lD);
		\draw[sync] (lD) to [loop right]
			node[right] {$\styleSync{\send\ActDefence{}\_\ok}$} (lD);
		\draw[sync] (l'1) to [loop right]
			node[right] {$\styleSync{\send\ActDefence{}\_\nok}$} (l'1);
	\end{tikzpicture}
	\vspace{1mm}
	}
  \\
\hline

\parbox[c]{4cm}{\centering
  \vspace{1mm}

\begingroup
\def\nodesubtree{\node[isosceles triangle, isosceles triangle apex angle=70, shape border rotate=90, minimum size=6.5mm]}

\begin{tikzpicture}
	[every node/.style={ultra thick,draw=red,minimum size=6mm}]
	\normalsize
	\node[and gate US,point up,logic gate inputs=ni, seq=9pt] (A)
		{\rotatebox{-90}{\large$\ActAttack$}};
	\nodesubtree             at (-1,-1.36) (a) {$\actAttack$};
	\nodesubtree[draw=Green] at ( 1,-1.32) (d) {$\!\actDefence\!$};
	\draw (a.north) -- ([yshift=1.4mm]a.north) -| (A.input 1);
	\draw (d.north) -- ([yshift=1.5mm]d.north) -| (A.input 2);
\end{tikzpicture}
\endgroup
	\vspace{1mm}
	}
& \parbox[c]{10cm}{\centering
	\vspace{1mm}
	\begin{tikzpicture}
			[every state/.style={thick,draw,minimum size=6mm},
				>=stealth',node distance=1.8cm,->]
		\node[state,initial] (l0) {$\loc_0$};
		\node[state,right of=l0] (l1) {$\loc_1$};
		\node[state,below of=l1] (l'1) {$\loc'_1$};
		\node[state,right of=l1] (l2) {$\loc_2$};
		\node[state,right of=l2] (lA) {$\loc_A$};
		\draw[sync] (l0) --
			node[above] {$\styleSync{\receive\actAttack{}\_\ok}$} (l1);
		\draw[sync] (l1) --
			node[above] {$\styleSync{\receive\actDefence{}\_\nok}$} (l2);
		\draw[sync] (l0) -- node[above,sloped]
			{$\styleSync{\receive\actAttack{}\_\nok}$} (l'1);
		\draw[sync] (l1) to node[above,sloped]
			{$\styleSync{\receive\actDefence{}\_\ok}$} (l'1);
		\draw (l2) -- node[above] {$\ActAttack{}$} (lA);
		\draw[sync] (lA) to [loop right]
			node[right] {$\styleSync{\send\ActAttack{}\_\ok}$} (lA);
		\draw[sync] (l'1) to [loop right]
			node[right] {$\styleSync{\send\ActAttack{}\_\nok}$} (l'1);
	\end{tikzpicture}
	\vspace{1mm}
	}
  \\
\hline

\parbox[c]{4cm}{\centering
  \vspace{1mm}

\begingroup
\def\nodesubtree{\node[isosceles triangle, isosceles triangle apex angle=70, shape border rotate=90, minimum size=6.5mm]}

\begin{tikzpicture}
	[every node/.style={ultra thick,draw=Green,minimum size=6mm}]
	\normalsize
	\node[and gate US,point up,logic gate inputs=ni, seq=9pt] (D)
		{\rotatebox{-90}{\large$\ActDefence$}};
	\nodesubtree           at (-1,-1.32) (d) {$\!\actDefence\!$};
	\nodesubtree[draw=red] at ( 1,-1.36) (a) {$\actAttack$};
	\draw (d.north) -- ([yshift=1.5mm]d.north) -| (D.input 1);
	\draw (a.north) -- ([yshift=1.4mm]a.north) -| (D.input 2);
\end{tikzpicture}
\endgroup
  \vspace{1mm}
	}
& \parbox[c]{10cm}{\centering
	\vspace{1mm}
	\begin{tikzpicture}
			[every state/.style={thick,draw,minimum size=6mm},
				>=stealth',node distance=1.8cm,->]
		\node[state,initial] (l0) {$\loc_0$};
		\node[state,right of=l0] (l1) {$\loc_1$};
		\node[state,below of=l1] (l'1) {$\loc'_1$};
		\node[state,right of=l1] (l2) {$\loc_2$};
		\node[state,right of=l2] (lD) {$\loc_D$};
		\draw[sync] (l0) --
			node[above] {$\styleSync{\receive\actDefence{}\_\ok}$} (l1);
		\draw[sync] (l1) --
			node[above] {$\styleSync{\receive\actAttack{}\_\nok}$} (l2);
		\draw[sync] (l0) -- node[above,sloped]
			{$\styleSync{\receive\actDefence{}\_\nok}$} (l'1);
		\draw[sync] (l1) to node[above,sloped]
			{$\styleSync{\receive\actAttack{}\_\ok}$} (l'1);
		\draw (l2) -- node[above] {$\ActDefence{}$} (l3);
		\draw[sync] (lD) to [loop right]
			node[right] {$\styleSync{\send\ActDefence{}\_\ok}$} (lD);
		\draw[sync] (l'1) to [loop right]
			node[right] {$\styleSync{\send\ActDefence{}\_\nok}$} (l'1);
	\end{tikzpicture}
	\vspace{1mm}
	}
  \\
\hline\hline
\end{tabular}
}
\end{center}
	\caption{\ADT/ sequential nodes and corresponding agent models\label{tab:transform:seq}}
\end{table}
}

\ifdefined\VersionLong
  \subsection{Correctness of the transformation}
  \label{sec:transformation:correctness}
\else
  \paragraph{Correctness of the transformation.}
\fi
\lp{The proof can be done for some essential patterns, and the proof for the others
will use exactly the same approach.}
Communication between two patterns involves pairwise synchronisation, given by
\cref{def:EIIS:gtrans:sync} in \Cref{def:EIIS}: the parent receives the status
from each child, updates its own status, and then informs its own parent.
Self-loops ensure that all parents are informed in DAG \ADT/s.  The order of
reception is not important as long as there is a representative sequence for the
attack/defence semantics, which the patterns ensure.
\LongVersion{\par}%
This covers \emph{expected} action signals, \ie those that a parent
must receive from its children to achieve its goal. For sequences containing
\emph{unexpected} signals it could happen that a parent does not receive all
information from its children, thus entering a deadlock before reaching its
final state to signal success to its own parent(s). Notice that these sequences
are precisely those that inhibit the goal of the (parent) gate. Therefore, not
reaching the final state is indeed the desired behaviour.

\ifdefined\VersionLong
  \subsection{Adding node attributes}
  \label{sec:transformation:attributes}
\else
  \paragraph{Adding node attributes.}
\fi
Transformation patterns in
\Cref{\LongVersion{sec:transformation:patterns}\ShortVersion{tab:transform:short}}
concern only an \ADT/ structure. To take attributes values into account, a child
must transmit these to all parents during transition synchronisation.
Attributes in a node are given by the \uline{intrinsic value} and
\uuline{computation function} of the node---see \Cref{sec:adt:attributes} and
the \EAMAS/ modification function in \cref{def:EAMAS:modfun} of
\Cref{def:EAMAS}. For example, the cost of an \gateAND gate is typically
\uline{its own cost}\uuline{~plus those of all children}.
Attributes values are computed \emph{after} all the node preconditions are satisfied%
\LongVersion{%
\ie in general when the action actually takes place (black transition arrows
in the automata)
}.
\gateOR gates involve a choice among children: here, computation happens
upon receiving the message from one child.
Conditions associated to \ADT/ countering nodes (\eg \texttt{TS} in
\Cref{fig:adt:treasure}) are added as guards to the corresponding action
transition.
\LongVersion{\par}%
Computation functions may be sensitive to the distribution of agents over nodes
for certain gates and attributes. For instance, time computation for \gateAND
gates is typically affected by the parallel operation of multiple agents with
the same goal; but this is not true in general for cost, neither for time in
\gateSAND gates. We illustrate this with our running example%
\footnote{%
	Computation functions for agents coalitions are further discussed in
	\Cref{sec:compfuns}.
}%
\!\!.

\ifdefined\VersionLong
  \subsection{Example: \EAMAS/ of the Treasure hunters}
  \label{sec:transformation:example}
\else
  \paragraph{Example: \EAMAS/ of the Treasure hunters.}
\fi
Applying the transformation patterns to the Treasure hunters \ADT/ of
\Cref{fig:adt:treasure} gives the \EAMAS/ in \Cref{fig:transformation:example}.
%
\begin{figure}[!ht]
	\vspace{-3ex}
	\centering
	\scalebox{0.8}{
	\begin{tikzpicture}[every state/.style={thick,draw,minimum size=6mm},
			>=stealth',node distance=1.8cm,->]
		\leafAgent[(0,0)]{f}{c_f:=100}{t_f:=120}
		\leafAgent[(5,0)]{b}{c_b:=500}{t_b:=60}
		\leafAgent[(10,0)]{h}{c_h:=500}{t_h:=3}
		\leafAgent[(0,-3)]{e}{}{t_e:=10}
		\leafAgent[(5,-3)]{p}{c_p:=100}{t_p:=10}
		\andAgent[(0,-6)]{\mathit{ST}}{b}{f}
			{c_\mathit{ST}:=c_f+c_b}{t_\mathit{ST}:=2+\mathit{timeC}(t_f,t_b)}
		\orAgent[(0,-9.8)]{\mathit{GA}}{h}{e}
			{}{c_\mathit{GA}:=c_h, t_\mathit{GA}:=t_h}
			{}{c_\mathit{GA}:=c_e, t_\mathit{GA}:=t_e}
		\sandAgent[(7.5,-6)]{\mathit{TF}}{\mathit{ST}}{\mathit{GA}}
			{c_\mathit{TF}:=c_\mathit{ST}+c_\mathit{GA}}
			{t_\mathit{TF}:=t_\mathit{ST}+t_\mathit{GA}}
		\nandAgent[(7.5,-9.8)]{\mathit{TS}}{\mathit{TF}}{p}
			{c_\mathit{TS}:=c_\mathit{TF}}
			{t_\mathit{TS}:=t_\mathit{TF}}
			{10 > t_\mathit{GA} + 2}
	\end{tikzpicture}
	}
	\caption{\EAMAS/ network for the \ADT/ shown in~\Cref{fig:adt:treasure}.}
	\label{fig:transformation:example}
	\vspace{-.5ex}
\end{figure}
%
\LongVersion{\par}%
\textcolor{PineGreen}{Green} labels indicate cost and time of each operation.
The computation for the \gateOR gate ($\mathit{GA}$) depends on the child from
which the message is received.
The condition associated to $\mathit{TS}$ is depicted in \textcolor{RedOrange}
{orange}. It checks whether the attack succeeds, but does not modify its cost
nor its time%
\LongVersion{, which are the same as those of its children}.
Time computation for the \gateAND gate $\mathit{ST}$ uses function
$\mathit{timeC}$%
\LongVersion{. This function is intended to return}%
\ShortVersion{, which returns}
the sum of its arguments if the children are handled by the same agent (there is
one thief) and their maximum otherwise (two or more thieves collaborate).
\LongVersion{\par}%
%
%
Note that sequential gates like \gateSAND enforce the (sequentially) dependent
execution of its children regardless of agents distributions. This models
situations of logical dependencies based on the nature of the attack/defence
instead of the resources available. In our example the thieves must complete
stealing the treasure \emph{before} going away, \LongVersion{and this will not change}
irrespective of the number of thieves. Hence the time for the thieves to flee is
\begingroup
\def\greentime#1{\ensuremath{\textcolor{PineGreen}{\mathit{#1}}}\xspace}
\mbox{$\greentime{t_{TF}}=\greentime{t_{ST}}+\greentime{t_{GA}}$}.
\endgroup

\section{Experiments}
\label{sec:expe}

To demonstrate our approach we implement the transformation patterns in \uppaal
and \imitator. With these popular and versatile software tools for quantitative
verification, we analyse the following case studies from the literature.

\subsection{Case studies}
\label{sec:expe:casestudies}

\paragraph{Forestalling a software release (\csfs).}
\LongVersion{%
This example models an attack to the intellectual property of company
$C$, and
is based on real studies performed (and anonymised) in \cite{BLPSW06}. In the
model an unlawful competitor company $U$ steals a software asset to get the
``first to the market'' advantage.  We follow the modelling pattern of
\cite{KumarRS15} which is sensitive to the order of events: software extraction
from $C$ must take place \emph{before} $U$ builds it into his own product. We
also specify that $U$ must deploy to market before $C$, which can only take
place after $U$ integrates the stolen software into his product.  The full
attack-defence tree enriched with (counter and failed reactive) defences is
presented in \Cref{fig:csfs}.
}
\ShortVersion{%
This example, shown in \Cref{fig:csfs} and based on a real-world instance
\cite{BLPSW06}, models an attack to the intellectual property of a company $C$,
by an unlawful competitor company $U$ aiming at being ``first to the market.''
}%
\begin{figure}[!h]
  \vspace{-1ex}
  \hspace{-.5em}
  \begin{subfigure}[b]{.6\linewidth}
	\centering
	\scalebox{0.75}{
	\begin{tikzpicture}
		[every node/.style={ultra thick,draw=red,minimum size=6mm},
		node distance=1.5cm]

		\node[and gate US,point up,logic gate inputs=nnn, seq=8pt] (FS)
			{\rotatebox{-90}{\texttt{FS}}};

		\node[state, below = 3mm of FS.west] (icp) {\texttt{icp}};
		\draw (icp.north) -- (FS.input 2);

		\node[state, right of = icp] (dtm) {\texttt{dtm}};
		\draw (dtm.north) -- ([yshift=0.15cm]dtm.north) -| (FS.input 3);

		\node[or gate US,point up,logic gate inputs=nnn,above of = icp,
			xshift=-4mm] (SC)
			{\rotatebox{-90}{\texttt{SC}}};
		\draw (SC.east) -- ([yshift=0.15cm]SC.east) -| (FS.input 1);

		\node[and gate US,point up,logic gate inputs=ni, seq=10pt,
			left of = SC] (NAS) {\rotatebox{-90}{\texttt{NAS}}};
		\draw (NAS.east) -- (SC.input 2);

		\node[and gate US,point up,logic gate inputs=ni,
			below of = NAS, node distance=3cm]
			(PRS) {\rotatebox{-90}{\texttt{PRS}}};
		\draw (PRS.east) -- ([yshift=0.15cm]PRS.east) -| (SC.input 3);

		\node[and gate US,point up,logic gate inputs=nn, seq=5pt,
			above of = NAS, node distance=3cm]
			(BRB) {\rotatebox{-90}{\texttt{BRB}}};
		\draw (BRB.east) -- ([yshift=0.15cm]BRB.east) -| (SC.input 1);

		\node[state, below = 4mm of BRB.west, xshift=-0.9cm]
			(bp) {\texttt{bp}};
		\draw (bp.north) -- ([yshift=0.15cm]bp.north) -| (BRB.input 1);

		\node[state, below = 4mm of BRB.west, xshift=0.9cm]
			(psc) {\texttt{psc}};
		\draw (psc.north) -- ([yshift=0.15cm]psc.north) -| (BRB.input 2);

		\node[and gate US,point up,logic gate inputs=nnn, seq=8pt,
			left = 8mm of NAS.input 1,xshift=-4mm]
			(NA) {\rotatebox{-90}{\texttt{NA}}};
		\draw (NA.east) -- ([yshift=0.15cm]NA.east) -| (NAS.input 1);

		\node[state, below = 3mm of NA.input 3] (heb) {\texttt{heb}};
		\draw (heb.north) -- (NA.input 3);

		\node[state, left of = heb] (sb) {\texttt{sb}};
		\draw (sb.north) -- ([yshift=0.1cm]sb.north) -| (NA.input 2);

		\node[state, left of = sb] (hh) {\texttt{hh}};
		\draw (hh.north) -- ([yshift=0.18cm]hh.north) -| (NA.input 1);

		\node[rectangle,draw=Green,minimum size=8mm,
			below = 4mm of NAS.west, xshift=10mm]
			(id) {\texttt{id}};
		\draw (id.north) -- ([yshift=0.15cm]id.north) -| (NAS.input 2);

		\node[and gate US,point up,logic gate inputs=nnn, seq=8pt,
			left = 8mm of PRS.input 1,xshift=-4mm]
			(PR) {\rotatebox{-90}{\texttt{PR}}};
		\draw (PR.east) -- ([yshift=0.15cm]PR.east) -| (PRS.input 1);

		\node[state, below = 3mm of PR.west] (reb) {\texttt{reb}};
		\draw (reb.north) -- (PR.input 2);

		\node[state, right of = reb] (rfc) {\texttt{rfc}};
		\draw (rfc.north) -- ([yshift=0.15cm]rfc.north) -| (PR.input 3);

		\node[state, left of = reb] (hr) {\texttt{hr}};
		\draw (hr.north) -- ([yshift=0.15cm]hr.north) -| (PR.input 1);

		\node[rectangle,draw=Green,minimum size=8mm,
			below = 4mm of PRS.west, xshift=10mm]
			(scr) {\texttt{scr}};
		\draw (scr.north) -- ([yshift=0.15cm]scr.north) -| (PRS.input 2);

	\end{tikzpicture}
	}
	\caption{\ADT/}
	\label{fig:csfs:adt}
  \end{subfigure}
  \hspace{-.1em}
  \begin{subfigure}[b]{.33\linewidth}
  \centering
  \scalebox{0.8}{
	\begin{tabular}{l@{$\,:\;$}l@{~}l@{~}r}
	\multicolumn{2}{l}{\textbf{Name}} & \textbf{Cost} & \textbf{Time}\\
	\hline
	\texttt{bp}  & bribe programmer       & \EUR{2k}  & 15~d\\
	\texttt{psc} & progr.\ steals code    & \EUR{0}   &  7~d\\
	\texttt{hh}  & hire hacker            & \EUR{1k}  & 20~d\\
	\texttt{sb}  & system has a bug       & \EUR{0}   &  0~d\\
	\texttt{heb} & hacker exploits bug    & \EUR{0}   &  3~d\\
	\texttt{id}  & intrusion detection   & \EUR{200} &  1~d\\
	\texttt{hr}  & hire robber            & \EUR{4k}  & 10~d\\
	\texttt{reb} & rob.\ enters building  & \EUR{500} &  3~d\\
	\texttt{rfc} & rob.\ finds code       & \EUR{0}   &  0~d\\
	\texttt{scr} & secure coding rooms    & \EUR{5k}  &  0~d\\
	\texttt{icp} & integr.\ code in prod. & \EUR{2k}  & 15~d\\
	\texttt{dtm} & deploy to market       & \EUR{1k}  &  5~d\\
	\texttt{BRB} & bribe                  & \EUR{0}   &  3~d\\
	\texttt{NA}  & network attack         & \EUR{0}   &  1~d\\
	\texttt{NAS} & network attack succ.   & \EUR{0}   &  0~d\\
	\texttt{PR}  & physical robbery       & \EUR{0}   &  0~d\\
	\texttt{PRS} & physical robbery succ. & \EUR{0}   &  0~d\\
	\texttt{SC}  & steal code             & \EUR{0}   &  0~d\\
	\texttt{FS}  & forestalling of softw. & \EUR{0}   & 10~d\\
	\end{tabular}
	}
	\caption{Attributes of nodes}
	\label{fig:csfs:attributes}
  \end{subfigure}
  \caption{\ADT/ of the \csfs case study}
  \label{fig:csfs}
  \vspace{-1ex}
\end{figure}
\ShortVersion{%
Software extraction from $C$ must take place \emph{before} $U$ builds it into
its own product. Also, $U$ must deploy to market before $C$, which can only take
place after $U$ integrates the stolen software into its product.
}

\paragraph{Compromise IoT device (\csiot).}
This case study describes an attack to an Internet-of-Things (IoT) device
either via wireless or wired LAN. Once the attacker gains access to the private
network and has acquired the corresponding credentials, it can exploit a
software vulnerability in the IoT device to run a malicious script.
\LongVersion{%
This model was extended in \cite{attack_FASE18} from an original attack tree
presented in \cite{SL15}. We further enrich the modelling with the addition of
defence mechanisms, effectively transforming it into the attack-defence tree
shown in \Cref{fig:csiot}.
}
\ShortVersion{%
\cite{attack_FASE18} presents an attack tree for this example as an extension of
the one in \cite{SL15}. The \ADT/ of \Cref{fig:csiot} enriches that with
defence nodes.
}

\begin{figure}[!!htb]
  \vspace{-1ex}
  \hspace{-1.5em}
  \begin{subfigure}[b]{.6\linewidth}
    \centering
	\scalebox{0.75}{
	\begin{tikzpicture}
		[every node/.style={ultra thick,draw=red,minimum size=6mm},
		node distance=1.5cm]

		\node[and gate US,point up,logic gate inputs=nnn, seq=8pt] (CIoTD)
			{\rotatebox{-90}{\texttt{CIoTD}}};

		\node[state, below = 3mm of CIoTD.west] (esv) {\texttt{esv}};
		\draw (esv.north) -- (CIoTD.input 2);

		\node[state, right of = esv] (rms) {\texttt{rms}};
		\draw (rms.north) -- ([yshift=0.15cm]rms.north) -| (CIoTD.input 3);

		\node[and gate US,point up,logic gate inputs=ni, seq=12pt,
			above of = esv, xshift=-2mm] (GAPNS)
			{\rotatebox{-90}{\texttt{APNS}}};
		\draw (GAPNS.east) -- ([yshift=0.15cm]GAPNS.east) -| (CIoTD.input 1);

		\node[and gate US,point up,logic gate inputs=nn,
			left = 8mm of GAPNS.input 1,xshift=-4mm]
			(GAPN) {\rotatebox{-90}{\texttt{APN}}};
		\draw (GAPN.east) -- ([yshift=0.15cm]GAPN.east) -| (GAPNS.input 1);

		\node[or gate US,point up,logic gate inputs=nn,
			below = 9mm of GAPN.input 1,yshift=14mm] (CPN)
			{\rotatebox{-90}{\texttt{CPN}}};
		\draw (CPN.east) -- ([yshift=0.15cm]CPN.east) -| (GAPN.input 1);

		\node[and gate US,point up,logic gate inputs=nn, seq=5pt,
			below = 9mm of CPN.input 1, yshift=12mm] (AL)
			{\rotatebox{-90}{\texttt{AL}}};
		\draw (AL.east) -- ([yshift=0.15cm]AL.east) -| (CPN.input 1);

		\node[state, below = 4mm of AL.input 1, xshift=-8mm]
			(flp) {\texttt{flp}};
		\draw (flp.north) -- ([yshift=0.15cm]flp.north) -| (AL.input 1);

		\node[state, below = 4mm of AL.input 2] (sma) {\texttt{sma}};
		\draw (sma.north) -- ([yshift=0.15cm]sma.north) -| (AL.input 2);

		\node[and gate US,point up,logic gate inputs=nn, seq=5pt,
			below = 9mm of CPN.input 2, yshift=-4mm] (AW)
			{\rotatebox{-90}{\texttt{AW}}};
		\draw (AW.east) -- ([yshift=0.15cm]AW.east) -| (CPN.input 2);

		\node[state, below = 4mm of AW.input 1] (fw) {\texttt{fw}};
		\draw (fw.north) -- ([yshift=0.15cm]fw.north) -| (AW.input 1);

		\node[state, below = 4mm of AW.input 2, xshift=8mm]
			(bwk) {\texttt{bwk}};
		\draw (bwk.north) -- ([yshift=0.15cm]bwk.north) -| (AW.input 2);

		\node[and gate US,point up,logic gate inputs=ni,
			below = 8mm of GAPN.input 2, yshift=-6mm]
			(GVC) {\rotatebox{-90}{\texttt{GVC}}};
		\draw (GVC.east) -- ([yshift=0.15cm]GVC.east) -| (GAPN.input 2);

		\node[state, below = 4mm of GVC.input 1] (gc) {\texttt{gc}};
		\draw (gc.north) -- ([yshift=0.15cm]gc.north) -| (GVC.input 1);

		\node[rectangle,draw=Green,minimum size=8mm,
			below = 4mm of GVC.west, xshift=10mm]
			(tla) {\texttt{tla}};
		\draw (tla.north) -- ([yshift=0.15cm]tla.north) -| (GVC.input 2);

		\node[rectangle,draw=Green,minimum size=8mm,
			below = 4mm of GAPNS.west, xshift=10mm]
			(inc) {\texttt{inc}};
		\draw (inc.north) -- ([yshift=0.15cm]inc.north) -| (GAPNS.input 2);
	\end{tikzpicture}
	}
	\caption{\ADT/}
	\label{fig:csiot:adt}
  \end{subfigure}
  \hspace{-.7em}
  \begin{subfigure}[b]{.33\linewidth}
  \centering
  \scalebox{0.8}{
	\begin{tabular}{l@{$\,:\;$}l@{~}l@{~}r}
	\multicolumn{2}{l}{\textbf{Name}} & \textbf{Cost} & \textbf{Time}\\
	\hline
	\texttt{flp}   & find LAN port            & \EUR{10}  &  1~h\\
	\texttt{sma}   & spoof MAC address        & \EUR{50}  & 30~m\\
	\texttt{fw}    & find WLAN                & \EUR{10}  &  5~h\\
	\texttt{bwk}   & break WPA keys           & \EUR{100} &  2~h\\
	\texttt{gc}    & get credentials          & \EUR{100} & 10~h\\
	\texttt{tla}   & two-level authentic.     & \EUR{5}   &  1~m\\
	\texttt{inc}   & inform of new connect.   & \EUR{5}   &  1~m\\
	\texttt{esv}   & exploit soft.\ vulnerab. & \EUR{10}  &  1~h\\
	\texttt{rms}   & run malicious script     & \EUR{100} & 30~m\\
	\texttt{AL}    & access LAN               & \EUR{0}   &  0~h\\
	\texttt{AW}    & access WLAN              & \EUR{0}   &  0~h\\
	\texttt{CPN}   & connect to private net.  & \EUR{0}   &  0~h\\
	\texttt{GVC}   & get valid credentials    & \EUR{0}   &  0~h\\
	\texttt{APN}   & access private net.      & \EUR{0}   &  3~m\\
	\texttt{APNS}  & access private net.\ succ. & \EUR{0} &  1~m\\
	\texttt{CIoTD} & compromise IoT device    & \EUR{0}   &  0~h\\
	\end{tabular}
	}
	\caption{Attributes of nodes}
	\label{fig:csiot:attributes}
  \end{subfigure}
  \caption{\ADT/ of the \csiot case study}
  \label{fig:csiot}
  \vspace{-1ex}
\end{figure}

\paragraph{Obtain admin privileges (\csadmin).}
\LongVersion{%
This scenario models an attacker trying to gain administrative privileges on a
UNIX system CLI, which requires either physical access to an already logged-in
console (entering the computer centre or corrupting an operator) or remote
access via privilege escalation (attacking the SysAdmin). This case study is
well known in the literature \cite{Wei91,Edg07,JW08,KPCS13,KumarRS15} and unlike
the previous ones it shows a mostly branching structure: all gates but one are
disjunctions in the original tree from \cite{Wei91}.  We enrich this scenario
with the \gateSAND from \cite{KumarRS15}, and further add reactive defences for
some of the attacks. The full attack-defence tree is presented in
\Cref{fig:csadmin}.

\begin{figure}[!h]
  \vspace{-4ex}
  \centering
  \caption{\ADT/ of the \csadmin case study}
  \label{fig:csadmin}
  \begin{subfigure}[b]{\linewidth}
	\centering
	\scalebox{0.82}{
	\begin{tikzpicture}
		[every node/.style={ultra thick,draw=red,minimum size=6mm},
		node distance=1.5cm]

		\node[or gate US,point up,logic gate inputs=nn] (OAP)
			{\rotatebox{-90}{\texttt{OAP}}};

		\node[or gate US,point up,logic gate inputs=nn,
			below = 12mm of OAP.input 1,yshift=14mm] (ACLI)
			{\rotatebox{-90}{\texttt{ACLI}}};
		\draw (ACLI.east) -- ([yshift=0.15cm]ACLI.east) -| (OAP.input 1);

		\node[state, below = 12mm of ACLI.input 1, xshift=-8mm]
			(co) {\texttt{co}};
		\draw (co.north) -- ([yshift=0.15cm]co.north) -| (ACLI.input 1);

		\node[and gate US,point up,logic gate inputs=ni,
			below = 23mm of ACLI, yshift = 7.75mm] (ECCS)
			{\rotatebox{-90}{\texttt{ECCS}}};
		\draw (ECCS.east) -- (ACLI.input 2);

		\node[or gate US,point up,logic gate inputs=nn,
			below = 12mm of ECCS.input 1,yshift=14mm] (ECC)
			{\rotatebox{-90}{\texttt{ECC}}};
		\draw (ECC.east) -- ([yshift=0.15cm]ECC.east) -| (ECCS.input 1);

		\node[state, below = 5mm of ECC.input 1, xshift=-6mm]
			(bcc) {\texttt{bcc}};
		\draw (bcc.north) -- ([yshift=0.15cm]bcc.north) -| (ECC.input 1);

		\node[state, below = 5mm of ECC.input 2, xshift=6mm]
			(ccg) {\texttt{ccg}};
		\draw (ccg.north) -- ([yshift=0.15cm]ccg.north) -| (ECC.input 2);

		\node[rectangle,draw=Green,minimum size=8mm,
			below = 5.5mm of ECCS.west, xshift=1.5mm]
			(scr) {\texttt{scr}};
		\draw (scr.north) -- (ECCS.input 2);

		\node[or gate US,point up,logic gate inputs=nnnn,
			below = 15mm of OAP.input 2,yshift=-3mm] (GSAP)
			{\rotatebox{-90}{\texttt{GSAP}}};
		\draw (GSAP.east) -- ([yshift=0.15cm]GSAP.east) -| (OAP.input 2);

		\node[and gate US,point up,logic gate inputs=ni, seq=12pt,
			below = 13mm of GSAP.west, yshift = 8.5mm] (GAPS)
			{\rotatebox{-90}{\texttt{GAPS}}};
		\draw (GAPS.east) -- (GSAP.input 1);

		\node[and gate US,point up,logic gate inputs=nn, seq=6pt,
			below = 12mm of GAPS.input 1, yshift = 4mm] (GAP)
			{\rotatebox{-90}{\texttt{GAP}}};
		\draw (GAP.east) -- (GAPS.input 1);

		\node[state, below = 5mm of GAP.input 1, xshift=-6mm]
			(opf) {\texttt{opf}};
		\draw (opf.north) -- ([yshift=0.15cm]opf.north) -| (GAP.input 1);

		\node[state, below = 5mm of GAP.input 2, xshift=6mm]
			(fgp) {\texttt{fgp}};
		\draw (fgp.north) -- ([yshift=0.15cm]fgp.north) -| (GAP.input 2);

		\node[rectangle,draw=Green,minimum size=8mm,
			below = 5mm of GAPS.input 2, xshift = 8mm] (tla)
			{\texttt{tla}};
		\draw (tla.north) -- ([yshift=0.15cm]tla.north) -| (GAPS.input 2);

		\node[and gate US,point up,logic gate inputs=ni,
			below = 13mm of GSAP.west, yshift = -20mm] (LSAS)
			{\rotatebox{-90}{\texttt{LSAS}}};
		\draw (LSAS.east) -- ([yshift=0.12cm]LSAS.east) -| (GSAP.input 2);

		\node[and gate US,point up,logic gate inputs=nnn, seq=6pt,
			below = 13mm of LSAS.input 1, yshift = 5.5mm] (LSA)
			{\rotatebox{-90}{\texttt{LSA}}};
		\draw (LSA.east) -- (LSAS.input 1);

		\node[state, below = 3mm of LSA.input 1, xshift=-7mm]
			(bsa) {\texttt{bsa}};
		\draw (bsa.north) -- ([yshift=0.15cm]bsa.north) -| (LSA.input 1);

		\node[state, below = 3mm of LSA.input 2]
			(vsa) {\texttt{vsa}};
		\draw (vsa.north) -- (LSA.input 2);

		\node[state, below = 3mm of LSA.input 3, xshift=7mm]
			(sat) {\texttt{sat}};
		\draw (sat.north) -- ([yshift=0.15cm]sat.north) -| (LSA.input 3);

		\node[rectangle,draw=Green,minimum size=8mm,
			below = 5mm of LSAS.input 2, xshift = 8mm] (nv)
			{\texttt{nv}};
		\draw (nv.north) -- ([yshift=0.15cm]nv.north) -| (LSAS.input 2);

		\node[and gate US,point up,logic gate inputs=ni,
			below = 13mm of GSAP.west, yshift = -48.5mm] (TSA)
			{\rotatebox{-90}{\texttt{TSA}}};
		\draw (TSA.east) -- ([yshift=0.3cm]TSA.east) -| (GSAP.input 3);

		\node[state, below = 6mm of TSA.input 2, xshift = -10mm]
			(th) {\texttt{th}};
		\draw (th.north) -- ([yshift=0.15cm]th.north) -| (TSA.input 1);

		\node[or gate US,point up,logic gate inputs=nn,draw=Green,
			below = 13mm of TSA.input 2,yshift=4.15mm] (DTH)
			{\rotatebox{-90}{\texttt{DTH}}};
		\draw (DTH.east) --(TSA.input 2);

		\node[rectangle,draw=Green,minimum size=8mm,
			below = 4mm of DTH.input 1, xshift = -6mm] (wd)
			{\texttt{wd}};
		\draw (wd.north) -- ([yshift=0.15cm]wd.north) -| (DTH.input 1);

		\node[rectangle,draw=Green,minimum size=8mm,
			below = 4mm of DTH.input 2, xshift = 6mm] (efw)
			{\texttt{efw}};
		\draw (efw.north) -- ([yshift=0.15cm]efw.north) -| (DTH.input 2);

		\node[state, below = 7mm of GSAP.input 4, xshift=60mm]
			(csa) {\texttt{csa}};
		\draw (csa.north) -- ([yshift=0.375cm]csa.north) -| (GSAP.input 4);

	\end{tikzpicture}
	}
	\vspace{1ex}
	\caption{\ADT/}
	\label{fig:csadmin:adt}
  \end{subfigure}
  \vspace{-4ex}
\end{figure}

\begin{figure}[!ht]\ContinuedFloat
  \centering
  \begin{subfigure}[b]{\linewidth}
	\centering
	\begin{tabular}{l@{$\,:\;$}l@{~}l@{~}r}
	\multicolumn{2}{l}{\textbf{Name}} & \textbf{Cost} & \textbf{Time}\\
	\hline
	\texttt{co}   & corrupt operator        & \EUR{4k}  &  4~d\\
	\texttt{bcc}  & break-in comp.\ centre  & \EUR{6k}  &  2~d\\
	\texttt{ccg}  & c.c.\ guest unwatched   & \EUR{100} &  5~d\\
	\texttt{scr}  & secure coding rooms     & \EUR{5k}  &  0~d\\
	\texttt{opf}  & obtain password file    & \EUR{100} &  3~d\\
	\texttt{fgp}  & find guessable pass.   & \EUR{0}   &  1~d\\
	\texttt{tla}  & two-level authentic.    & \EUR{5}   &  1~m\\
	\texttt{bsa}  & befriend Sys.\ Admin.   & \EUR{500} & 14~d\\
	\texttt{vsa}  & visit SA at work        & \EUR{20}  &  2~d\\
	\texttt{sat}  & spy SA terminal         & \EUR{0}   & 30~m\\
	\texttt{nv}   & no-visits policy        & \EUR{0}   &  0~d\\
	\texttt{th}   & trojan horse SA         & \EUR{100} &  3~d\\
	\texttt{wd}   & watchdog sys.\ daemon   & \EUR{2k}  &  5~m\\
	\texttt{efw}  & E-Mail firewall         & \EUR{3k}  &  0~m\\
	\texttt{csa}  & corrupt Sys.\ Admin.    & \EUR{5k}  &  5~d\\
	\texttt{ECC}  & enter computer centre   & \EUR{0}   &  0~d\\
	\texttt{ECCS} & enter c.c.\ successful  & \EUR{0}   &  1~h\\
	\texttt{GAP}  & get admin password      & \EUR{0}   & 10~m\\
	\texttt{GAPS} & GAP successful          & \EUR{0}   &  2~m\\
	\texttt{LSA}  & look over SA shoulder   & \EUR{0}   &  0~m\\
	\texttt{LSAS} & LSA successful          & \EUR{0}   &  0~m\\
	\texttt{DTH}  & defence against trojans & \EUR{0}   &  0~m\\
	\texttt{TSA}  & trojan horse for SA     & \EUR{0}   &  0~m\\
	\texttt{ACLI} & access c.c.\ CLI        & \EUR{0}   &  2~m\\
	\texttt{GSAP} & get SA password         & \EUR{0}   &  0~m\\
	\texttt{OAP}  & obtain admin privileges & \EUR{0}   &  0~m\\
	\end{tabular}
	\vspace{2ex}
	\caption{Attributes of nodes}
	\label{fig:csadmin:attributes}
  \end{subfigure}
\end{figure}

}
\ShortVersion{%
To gain administrative privileges on a UNIX system, an attacker needs either
physical access to an already logged-in console (entering the computer centre or
corrupting an operator) or remote access via privilege escalation (attacking the
SysAdmin). This well-known case study \cite{Wei91,Edg07,JW08,KPCS13,KumarRS15}
exhibits a mostly branching structure: all gates but one are disjunctions in
the
original tree from \cite{Wei91}. We enrich this scenario with the \gateSAND from
\cite{KumarRS15}, and add reactive defences. The full \ADT/ is presented in
\Cref{fig:csadmin} of \Cref{sec:gainadmin}.
}

\subsection{Uppaal}
\label{sec:expe:uppaal}

All case studies from \Cref{sec:expe:casestudies} were modelled in \uppaal
\cite{BDL04,DLLMP15}\LongVersion{. \uppaal is}\ShortVersion{,} a state-of-the-art
tool for validation and
verification of real-time systems\LongVersion{, developed by the universities of Uppsala
and
Aalborg}.  Automata templates are declared via its \LongVersion{graphical user interface}\ShortVersion
{GUI} and
later instantiated for model verification via PCTL-like queries. This allows a
straightforward encoding of the \ADT/-to-\EAMAS/ translations\LongVersion{ (\ie the
agent
models in \crefrange{tab:transform:leaf}{tab:transform:seq})} as \uppaal
templates. \LongVersion{Then m}\ShortVersion{M}odelling an \ADT/ is \LongVersion{just
a matter of}\ShortVersion{achieved by} instantiating these
templates\LongVersion{ following the structure of the concrete \ADT/}. Moreover, support
for
agents distribution can be easily encoded \LongVersion{using immutable arrays: the user
can
thus assign arbitrary}\ShortVersion{assigning} agents to tree nodes \LongVersion{by
modifying an array specifically
defined for that purpose}\ShortVersion{in a dedicated array}.

We compared the original \uppaal models from \cite{KumarRS15,attack_FASE18}
(extended with defences as detailed in \Cref{sec:expe:casestudies}) with our
\EAMAS/ translations. Our goal was twofold: on the one hand we verified
correctness, checking that reachability queries (``can the attack succeed?'')
and quantitative queries (``what is the time of the fastest possible attack?'',
``what is the cost incurred?'') coincide between our models and those from
\cite{KumarRS15,attack_FASE18}; on the other hand (for the \EAMAS/ translations)
we studied the impact of agent coalitions on these metrics.

\uppaal modelling in \cite{KumarRS15,attack_FASE18} uses clock variables to
represent time, and clock constraints to encode the duration of attacks and
defences. This is the natural approach in a tool designed for the verification
of timed systems, yet it has two drawbacks when it comes to the analysis of
\ADT/s:
\begin{itemize}
	\item	\LongVersion{When clocks are defined and the model is recognised as a
	timed system, }\uppaal uses abstractions and approximations for time regions that rule out decidability in the general case \cite{BDL04}. Thus reachability queries like ``can the attack be performed'' can at best result in a ``this may be true'' answer.
	\LongVersion{\begin{itemize}
	\item	In contrast, \EAMAS/ transformations are untimed (see \Cref{sec:ADT,sec:EAMAS}) and verification is exact. The price to pay is a larger state space and approximately $3\times$ longer computation times (depending on the model), at least in \uppaal academic v.~4.1.19.
	\end{itemize}}%
	\ShortVersion{
	In contrast, \EAMAS/ transformations are untimed (see \Cref{sec:ADT,sec:EAMAS}) and verification is exact. The price to pay is a larger state space and approximately $3\times$ longer computation times.
	}
	\item	Unless explicitly encoded case-per-case in the \LongVersion{(templates
	of the) }gates, time elapses simultaneously for all clocks. Thus, modelling
	the dependent execution of nodes \LongVersion{(for instance to signify that the
	same person must perform all attacks/defences) }requires \eg using a \gateSAND
	rather than an \gateAND. This contaminates the structure of the \ADT/
	with the distribution of the agents\LongVersion{ performing it} for analysis.
	\LongVersion{%
	\begin{itemize}
	\item	\EAMAS/ was designed with the study of agents coalitions
	in mind, and for the \ADT/ transformations
	this implies a clear separation between the tree structure and
	the agents performing it. This impacts the quantitative analyses as shown next.
	\end{itemize}
	}%
	\ShortVersion{%
	\EAMAS/ can keep the \ADT/ structure unmodified while studying agents
	coalitions.
	}
\end{itemize}

\paragraph{Case study \csfs.}
The minimum time for a successful attack is 43~days, via the \gateNAND attack
gate \texttt{PRS}.  Both \uppaal models of the system (\EAMAS/ and
\cite{KumarRS15}) produced the same correct result.
The maximum time for a successful attack is 92 days: this implies a failed
network attack (\texttt{NA}) due to the presence of the intrusion detection
(\texttt{id}) defence, a failed physical robbery (\texttt{PR}) due to the
company's use of secure coding rooms (\texttt{scr}), and a successful
bribery to a programmer (\texttt{BRB}).  Notice that the time for \texttt{NA},
\texttt{PR}, and \texttt{BR} \emph{must be added} if we consider the same
person is performing all actions, \ie there is only one attacker. This cannot
be enforced in the \uppaal model from \cite{KumarRS15} without sequential
\gateOR gates, in the absence of which the query for maximum time yields ``55
days''\LongVersion{: 15+7+3 days to perform \texttt{BRB} (the longest attack) and
15+5+10
days to perform the rest of the \gateSAND gate \texttt{FS}}.  The \EAMAS/
\uppaal model can produce the correct result (92 days) by specifying that a
single agent is performing all attacks; and it also yields ``55 days'' if
independent agents are used for the \texttt{BRB}, \texttt{NA}, and \texttt{PR}
attacks.
In the case of costs, the cheapest attack is performed via
\texttt{NAS} (\EUR{4k}) whereas the most expensive
one (\EUR{10500}) involves all attacks, with \LongVersion{corresponding }failures of
\texttt{NA} and \texttt{PR} due to the \texttt{id} and \texttt{scr} defences.
Both models yield the correct values to these queries.

\paragraph{Case study \csiot.}
In this case the same subtree is the source of the fastest and cheapest attack,
via the access LAN strategy (\texttt{AL}) to connect to the private network
(\texttt{CPN}).
The minimum cost is \EUR{270} and the maximum one is \EUR{380}, when
\texttt{AW} is chosen in addition to \texttt{AL}%
\LongVersion{\footnote{One could also compute the most expensive \emph{rational}
	attack, where either \texttt{AW} or \texttt{AL} is performed,
	which results in \EUR{320}.}}.
The \EAMAS/ and the \cite{attack_FASE18} \uppaal models yield correct results.
For this \ADT/ both the minimum and maximum time depend on the agent coalition.
This is because the access private network subtree (\texttt{APN}) is an
\gateAND gate, which can be executed in parallel or not depending on how many
attackers are involved.
For this reason and given its use of clocks, the \cite{attack_FASE18} model can
only produce 694~minutes as the fastest attack, which is correct (and the
\EAMAS/ model also computes) when different agents perform the \texttt{CPN} and
\texttt{GVC} attacks. When instead the same agent is used for both attacks, the
\EAMAS/ models returns the correct 784~minutes minimum time. The only way to
achieve this in the \cite{attack_FASE18} model is to modify the tree, replacing
the \gateAND for a \gateSAND in \texttt{APN}. Furthermore\LongVersion{ and following
the
same logic than for the \csfs case study}, the longest attack takes
1204~minutes, which can be computed from the \ADT/ of \Cref{fig:csiot} by the
\EAMAS/ model but not by the \cite{attack_FASE18} model without modifications to the
tree structure\LongVersion{, namely the \gateOR gate \texttt{CPN}}.

\paragraph{Case study \csadmin.}
The fastest attack takes 2942~minutes\LongVersion{ ($\approx2$~days)} via breaking
into the
computer centre to access a connected CLI (\texttt{ACLI} through \texttt{bcc}),
whereas the slowest takes 23070~minutes\LongVersion{ ($\approx16$~days)} via looking
over
the SA's shoulder (\texttt{LSAS}). For full parallelism of attacks, these
values are computed correctly by both the \cite{KumarRS15} and the \EAMAS/
models. To study other agent coalitions, \cite{KumarRS15} suffers from the same
limitations as in the other two case studies. \LongVersion{Notice that agent assignment
is irrelevant for minimisations in this \ADT/ since every gate is either \gateSAND
or \gateOR.}
Regarding costs, there are two cheapest attacks (\EUR{100}) via either
\texttt{GAPS} or \texttt{TSA}. The \LongVersion{most expensive rational
attack (\EUR{6k}) involves breaking into the computer centre (\texttt{bcc}),
the absolute }most expensive attack (\EUR{15820}) is the summation of the
costs of all attack leaves.  Again both values are correctly computed by the
two models.
\LongVersion{Interestingly, the}\ShortVersion{The} reachability query ``is an attack
feasible?''is computed \LongVersion{in
a standard laptop} for the \EAMAS/ model in a few milliseconds, whereas it
takes more than three minutes for the \cite{KumarRS15} model. This is a
consequence of the reduction techniques applied to timed-systems\LongVersion{: time
constraints on clocks create an intricate time-region graph} where reachability
is nontrivial; in contrast, the full state space exploration in the \EAMAS/
model can quickly reach \LongVersion{(via depth- or breadth-first search) }the goal
state starting from the \texttt{csa} or \texttt{co} leaves. \LongVersion{Oppositely
and as}\ShortVersion{As}
discussed earlier, state space bloating in the \EAMAS/ model impacts
quantitative queries.


\subsection{Imitator}
\label{sec:expe:imitator}

\LongVersion{%
\imitator{} \cite{AndreFKS12} is a model-checker for Parametric Timed Automata
\cite{AlurHV93}.
Models are composed of several automata which synchronise on transitions. All
automata declaring the synchronisation must participate together. Thus, the
message passing between a child and its only parent in an \ADT/ is captured
by a synchronisation transition between the two of them (for DAGs, all
possibilities must be explicitly modelled).

We implemented our running example in \imitator{} (\url{www.imitator.fr}), as
the automata of \Cref{fig:transformation:example}. The fixed cost and time
values in the table of \Cref{fig:ths:attributes} are declared as constants,
while the current cost and time of each \ADT/ node is a variable.
The property of interest is the reachability of location $l_\mathit{TS}$. It could
be checked by searching for the existence of such a state, but also by generating
and exploring the full state space, thus getting all possible cost/time combinations
for the solutions.
The minimal and maximal costs and times of the other case studies were also found
by state space exploration with \imitator{}, confirming the results obtained with
\uppaal.

More interestingly, we experimented with the parametric capabilities of \imitator{}.
In the treasure hunters example, we notice that the thief has a solution to flee.
Hence, we considered the question: ``What is the maximum time for the police to arrive
early enough to catch
the thieves?''. To address this question, the time for \ADT/ node \texttt{p} is no more
a constant but a parameter. The synthesis algorithm of \imitator{} then returns a constraint
on parameters and constants that states the police should take at most 5 minutes.

For the other case studies, we considered an attack to be successful if its associated
defence was slower.

For \csfs, we studied the \texttt{id} defence and proved that it should take at most
than 1 min for blocking attacks. Similarly, in \csiot, the \texttt{inc} defence
is effective iff it takes at most 3 minutes. Finally, we proved that whatever the
time for \texttt{tla} in the \csadmin{} case study, an attack can be achieved. Hence,
the other defences are required.
}
\ShortVersion{%
\imitator{} (\url{www.imitator.fr}, \cite{AndreFKS12}) is a model-checker for
Parametric Timed Automata \cite{AlurHV93}.  Models are composed of automata that
synchronise on shared transitions (actions); in \ADT/s this involves message
passing between a child and its parent(s), where all combinations must be
explicit for DAGs.

We modelled our running example in \imitator{} as the \EIIS/ of
\Cref{fig:transformation:example}. Cost and time are constants for the nodes
with fixed attributes---see the table in \Cref{fig:ths:attributes}---and
variables for the remaining nodes. We could verify that location $l_\mathit{TS}$
is reachable in the model, \ie \emph{the attack is feasible for the selected
attributes values.} This holds even for a single attacker, since \texttt{GA}
must be done after \texttt{ST}, and the police is alerted as soon as \texttt{b}
and \texttt{f} succeed (but before \texttt{ST} does).

Moreover, we considered the question ``To catch the thieves, what is the maximum
time the police can take to arrive?'' This requires synthesising a value for the
time attribute of the \texttt{p} node, which becomes a model \emph{parameter}.
\imitator{} computed that the police can take at most 5~minutes to prevent the
burglary.

For the three case studies of \Cref{sec:expe:casestudies}, the min/max costs and
times were also found by state space exploration with \imitator{}, confirming
the results obtained with \uppaal.
Regarding parameter synthesis, we considered an attack is successful if its
associated defence was slower:
\begin{enumerate*}[($i$)]
	\item for \csfs, we computed that \texttt{id} should take at most 1~day to block
		  \texttt{NA}---since \texttt{NAS} is a \emph{failed reactive defence},
		  \texttt{id} is triggered as soon as \texttt{heb} succeeds, and has to
	      finish faster than the intrinsic time of \texttt{NA};
	\item for \csiot, \texttt{inc} is effective iff it takes at most 3~minutes;
	\item finally, for \csadmin we proved that whatever the time for \texttt{tla},
	      an attack is feasible (as \texttt{GSAP} is a disjunction), and hence
	      the other defences are required.
\end{enumerate*}
}


\LongVersion{
\subsection{Verics}
\label{sec:expe:verics}

\subsection{PTA2SMT}
\label{sec:expe:pta2smt}
}

\section{Conclusion and Future work}
\label{sec:conclu}

\ceb{new \ADT/ construct}
In this work we revisited attack-defence trees under a unified syntax, extending
the usual constructs with a new sequential counter-operator. Our syntax framed
this extension naturally among the other constructs, and in accordance to the
chosen semantics and other related work.
\ceb{new \EAMAS/ + approach by patterns + compositionality + reduced state space}
More importantly, we extended \AMAS/ to model \ADT/s in an agent-aware
formalism---\EAMAS/---and provided transformation patterns from \ADT/ constructs
to agent models. Our \ADT/--\EAMAS/ transformation preserves the compositional
system description, ensuring a succinct state space where reachability queries
can be efficiently checked.
\ceb{flexibility for attributes + agents configurations}
\EAMAS/ models decorate all \ADT/ nodes with attributes and an operating agent.
We exploited this flexibility to analyse (via \uppaal on three case studies) the
impact of different agent coalitions on attack/defence performance metrics such
as time and cost.
\ceb{parameter synthesis}
Finally, we analised the feasibility of specific goals using \imitator: we
synthesised parameters (\ie attribute values of \ADT/ nodes) rendering an
attack/defence feasible, or we proved them infeasible.
\ceb{encouraging experiments}
Overall, our experimental results show how different agent distributions affect
the time of attacks/defence strategies, possibly rendering some infeasible. We
expect this will open the gate to richer studies of real-world security
scenarios, with multiple distinct agents.

Our next objectives include designing logics to express properties of \EAMAS/,
and adapt the \emph{partial order reduction} (POR) methodology presented
in~\cite{AAMASWJWPPDAM2018a} to reason about the abilities of agents in \EAMAS/
models. Moreover, we will extend this to handling parametric timing information.
Finally, we plan to extend our tool that currently transforms \ADT/s into
\EAMAS/ models, to automatically generate \imitator{} and \uppaal{} models.


\newcommand{\etalchar}[1]{$^{#1}$}

\appendix

\section{Computation functions for \EAMAS/ agents}
\label{sec:compfuns}

\begingroup

\def\node#1{\ensuremath{\mathtt{#1}}\xspace}
\def\fun#1{\ensuremath{\mathit{#1}}\xspace}
\def\initime#1{\fun{init\_time}({#1})}
\def\greentime#1{\ensuremath{\textcolor{PineGreen}{\mathit{#1}}}\xspace}

To model and study the impact of different strategies and agent coalitions on
\ADT/s, computation functions for node attributes are a versatile solution.
In our running example, the computation function for the time of the attack gate
\mbox{$\gate{ST}=\gateAND(\node{b},\node{f})$} can be generically expressed with
functions $f$ and $g$:
\[
	f\left( \initime{\gate{ST}},g(\initime{\node{b}},\initime{\node{f}}) \right)
\]
\begin{itemize}[label=$\blacktriangleright$,topsep=.5ex,itemsep=.5ex]
\item	$f=+$ necessarily, because the operations involved in stealing
		the treasure depend logically on the completion of both
		children nodes, and
\item	$g$ depends on the agent coalition: if two or more thieves cooperate
		then \mbox{$g=\max$;} but if there is a single attacker then $g=+$.
\end{itemize}

\begin{figure}[ht]
  \vspace{-2.5ex}
  \caption{Running example: the treasure hunters}
  \begin{subfigure}[b]{.45\linewidth}
	\centering
	\scalebox{.85}{
	\begin{tikzpicture}
		[every node/.style={ultra thick,draw=red,minimum size=6mm}]
		\node[and gate US,point up,logic gate inputs=ni] (ca)
		{\rotatebox{-90}{\texttt{TS}}};
		\node[rectangle,draw=Green,minimum size=8mm, below = 5mm of ca.west, xshift=10mm] (d) {\texttt{p}};
		\draw (d.north) -- ([yshift=0.28cm]d.north) -| (ca.input 2);
		\node[and gate US,point up,logic gate inputs=nn, seq=4pt, below = 9mm of ca.west, yshift=14mm] (A)
		{\rotatebox{-90}{\texttt{TF}}};
		\draw (A.east) -- ([yshift=0.15cm]A.east) -| (ca.input 1);
		\node[and gate US,point up,logic gate inputs=nn, below = 9mm of A.west, yshift=14mm] (a1)
		{\rotatebox{-90}{\texttt{ST}}};
		\draw (a1.east) -- ([yshift=0.15cm]a1.east) -| (A.input 1);
		\node[state, below = 4mm of a1.west, xshift=-5mm] (a1n) {\texttt{b}};
		\draw (a1n.north) -- ([yshift=0.15cm]a1n.north) -| (a1.input 1);
		\node[state, below=4mm of a1.west, xshift=5mm] (a11) {\texttt{f}};
		\draw (a11.north) -- ([yshift=0.15cm]a11.north) -| (a1.input 2);
		\node[or gate US,point up,logic gate inputs=nn, below = 10mm of A.west, yshift=-7mm] (a2)
		{\rotatebox{-90}{\texttt{GA}}};
		\draw (a2.east) -- ([yshift=0.15cm]a2.east) -| (A.input 2);
		\node[state, below = 4mm of a2.west, xshift=-5mm] (a21) {\texttt{h}};
		\draw (a21.north) -- ([yshift=0.15cm]a21.north) -| (a2.input 1);
		\node[state, below=4mm of a2.west, xshift=5mm] (a2n) {\texttt{e}};
		\draw (a2n.north) -- ([yshift=0.15cm]a2n.north) -| (a2.input 2);
	  \end{tikzpicture}
	}
	\vspace{1ex}
	\caption{\ADT/}
  \end{subfigure}
  \hfill
  \begin{subfigure}[b]{.45\linewidth}
	\centering
	\scalebox{.9}{\parbox{\linewidth}{%
		\begin{tabular}{r@{~}l@{~\;}r@{~\;}r}
			\multicolumn{2}{l}{\bf Name} & {\bf Cost} & {\bf Time} \\
			\hline
			\texttt{TS} & (treasure stolen)  & &\\
			\texttt{p}  & (police)           & 100\,\euro{} & 10 min\\
			\texttt{TF} & (thieves fleeing)  & & \\
			\texttt{ST} & (steal treasure)   &              & 2 min \\
			\texttt{b}  & (bribe gatekeeper) & 500\,\euro{} & 1 h \\
			\texttt{f}  & (force arm. door)  & 100\,\euro{} & 2 h \\
			\texttt{GA} & (go away)          & & \\
			\texttt{h}  & (helicopter)       & 500\,\euro{} & 3 min \\
			\texttt{e}  & (emergency exit)   &              & 10 min
		\end{tabular}
		~\\[2ex]
		\uline{\textbf{Condition for \texttt{TS}}}:\\[.5ex]
		\mbox{$\initTime{\mathtt{p}} > \Time{\mathtt{ST}} + \Time{\mathtt{GA}}$}
	  }}
	  \vspace{2ex}
	  \caption{Attributes of nodes}
  \end{subfigure}
  \vspace{-2.5ex}
\end{figure}

In the general case, the children of a gate will be subtrees \node{L} and
\node{R} rather than leaves, which allows more complex computations of potential
parallelism between the agents in \node{L} and \node{R}.  One could for instance
encode a simple worst-case scenario: take an \gateAND gate involving agents
$\mathit{A}$, and let $\mathit{A_\node{L}}\subseteq\mathit{A}$ (resp.\
$\mathit{A_\node{R}}\subseteq\mathit{A}$) be all agents from subtree \node{L}
(resp.\ \node{R}). Then make the time of the gate be either%
\begin{itemize}[label=$\blacktriangleright$,topsep=.5ex,itemsep=.5ex]
\item	\emph{the sum} of the times taken to complete \node{L} and \node{R} if
		$\mathit{A_\node{L}}\cap\mathit{A_\node{R}}\neq\varnothing$, because
		then some agents are present in both children of the \gateAND, or
\item	\emph{the maximum} between these times otherwise, because then \node{L}
		and \node{R} are fully independent.
\end{itemize}

Furthermore, the sharing of children nodes by different gates in DAG \ADT/s can
also be captured in any desired way, by a careful choice of the computation
functions for each attribute.  Take for instance the tree
\begin{equation}
	\gateAND(\gateSAND_1(\node{a_1,a_2}),\gateSAND_2(\node{a_2,a_3}))
	\label{eq:DAG_ADT}
\end{equation}
where $\node{a_1,a_2,a_3}$ are attack leaves and $\gateSAND_1,\gateSAND_2$ are
standard \gateSAND gates. Due to the sequential operation of these gates, the
computation function for the time attribute of $\gateSAND_1$ must be
$\fun{time}(\gateSAND_1)=\initime{\gateSAND_1}+\initime{\node{a_1}}+\initime{\node{a_2}}$,
and analogously for $\gateSAND_2$.  However, the time of the full \gateAND
attack is
\begin{equation}
	\fun{time}(\gateAND) = 
		\sum_{i=1}^3\initime{\node{a_i}} + \sum_{j=1}^2 \initime{\gateSAND_j}.
	\label{eq:DAG_ADT_timefun}
\end{equation}
In particular, even for a single agent, $\initime{\node{a_2}}$ must be counted
only once: when $\gateSAND_1$ finishes, it means that \node{a_2} has already
been completed for $\gateSAND_2$ too.

To encode this compositionally via computation functions, one must operate at
the level of the parent(s) that close the DAG loop, \eg the \gateAND in
\cref{eq:DAG_ADT}. This is because only at those nodes the sharing of children 
\begin{enumerate*}[$(i)$]
\item	can be realised, and
\item	it may have an impact in computations
\end{enumerate*}:
$\fun{time}(\gateSAND_1)$ considered individually is not influenced by
operations in $\gateSAND_2$.

Let us finally analyse time computations for a coalition of two agents, \ie
two attackers cooperate. In this case, $\gateSAND_1$ and $\gateSAND_2$ can
\emph{neither} operate in parallel, because \node{a_2} is \emph{the same attack}
for both gates. The root cause of this restriction is that the tree in
\cref{eq:DAG_ADT} is isomorphic to $\gateSAND(\node{a_1,a_2,a_3})$, so
\cref{eq:DAG_ADT_timefun} is the only correct way to compute
$\fun{time}(\gateAND)$, regardless of agent coalitions.  Again, computation
functions can encode this by operating at the level of the \gateAND gate, where
the shared children in the DAG re-join.

\endgroup

\section{Case study: \csadmin}
\label{sec:gainadmin}

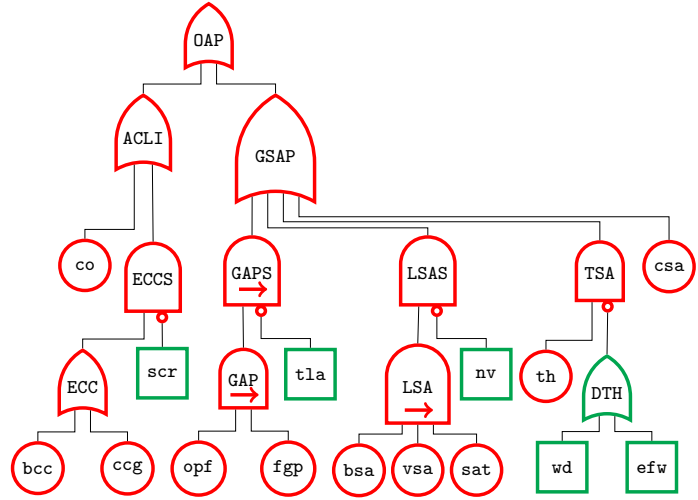
\end{document}